\documentclass[12pt]{article}
\usepackage{graphics,graphicx}
\usepackage{amssymb,epsfig,amsmath,euscript,array}
\usepackage{cite}

\usepackage{color}
\usepackage{hyperref}
\hypersetup{
    colorlinks,
    citecolor=black,
    filecolor=black,
    linkcolor=black,
    urlcolor=black
  }
  \hypersetup{linktocpage}

\makeatletter \@addtoreset{equation}{section} \makeatother

\def \shuffle{{\,\amalg\hskip -4.2pt\amalg\,}}
\newcommand \vev [1] {\langle\,{#1}\,\rangle}
\newcommand \vevv [1] {[\,{#1}\,]}

\newcounter{multieqs}

\newcommand{\be}{\begin{equation}}
\newcommand{\ee}{\end{equation}}

\newcommand{\bm}[1]{\mbox{\boldmath $#1$}}

\newcommand{\kslash}{k \!\!\! / }

\newcommand{\lslash}{l \!\! / }
\newcommand{\Pslash}{P \!\!\!\! / }

\newcommand{\islash}{i \!\!\! / }
\newcommand{\jslash}{j \!\!\! / }
\newcommand{\aslash}{a \!\!\! / }
\newcommand{\bslash}{{b \hspace{-6pt} \slash} }

\newcommand{\onslash}{1 \!\!\! / }
\newcommand{\twslash}{2 \!\!\!/ }
\newcommand{\thslash}{3 \!\!\!/ }
\newcommand{\foslash}{4 \!\!\! / }
\newcommand{\fislash}{5 \!\!\! / }

\newcommand{\mslash}{m \!\!\! / }

\def\bd{\begin{document}}
\def\ed{\end{document}}
\def\nn{\nonumber}
\def\bea{\begin{eqnarray}}
\def\eea{\end{eqnarray}}

\def\red{\color{red}}
\def\black{\color{black}}
\def\blue{\color{blue}}
\def\orange{\color{orange}}

\def\ab{(ijab)}
\def\ba{(ijba)}
\def\ijab{{\tr}_{+}(\islash\, \jslash\, \aslash \, \bslash)}
\def\ijba{{\tr}_{+}(\islash\, \jslash\, \bslash \, \aslash)}
\def\ijaP{{\tr}_{+}(\islash\, \jslash\, \aslash \, \Pslash)}
\def\ijPLa{{\tr}_{+}(\islash\, \jslash\, \Pslash_L \, \aslash)}
\def\ijaPL{{\tr}_{+}(\islash\, \jslash\, \aslash \, \Pslash_L)}
\def\ijPLza{{\tr}_{+}(\islash\, \jslash\, \Pslash_{L;z} \, \aslash)}
\def\ijaPLz{{\tr}_{+}(\islash\, \jslash\, \aslash \, \Pslash_{L;z})}
\def\ijPa{{\tr}_{+}(\islash\, \jslash\, \Pslash \, \aslash)}
\def\iaPb{{\tr}_{+}(\islash\, \aslash\, \Pslash \, \bslash)}
\def\ibPa{{\tr}_{+}(\islash\, \bslash\, \Pslash \, \aslash)}
\def\ijPmu{{\tr}_{+}(\islash\, \jslash\, \Pslash \, \mu)}
\def\ibmuP{{\tr}_{+}(\islash\, \bslash\, \mu \, \Pslash)}
\def\ibmua{{\tr}_{+}(\islash\, \bslash\, \mu \, \aslash)}
\def\iamub{{\tr}_{+}(\islash\, \aslash\, \mu \, \bslash)}
\def\jaPb{{\tr}_{+}(\jslash\, \aslash\, \Pslash \, \bslash)}
\def\ijmuP{{\tr}_{+}(\islash\, \jslash\, \mu \, \Pslash)}
\def\ijmum{{\tr}_{+}(\islash\, \jslash\, \mu \, \mslash)}
\def\ijmmu{{\tr}_{+}(\islash\, \jslash\, \mslash \, \mu)}
\def\ijmP{{\tr}_{+}(\islash\, \jslash\, \mslash \, \Pslash)}
\def\iabP{{\tr}_{+}(\islash\, \aslash\, \bslash \, \Pslash)}
\def\ijbP{{\tr}_{+}(\islash\, \jslash\, \bslash \, \Pslash)}
\def\jbPa{{\tr}_{+}(\jslash\, \bslash\, \Pslash \, \aslash)}
\def\ijPb{{\tr}_{+}(\islash\, \jslash\, \Pslash \, \bslash)}
\def\jbmua{{\tr}_{+}(\jslash\, \bslash\, \mu \, \aslash)}

\def\loablt{ {\tr}_{+}(\lslash_1\, \aslash \, \bslash\, \lslash_2)}

 \def\ijlolt{{\tr}_{+}(\islash\, \jslash\, \lslash_1 \, \lslash_2)}
\def\ijltlo{{\tr}_{+}(\islash\, \jslash\, \lslash_2 \, \lslash_1)}
\def\ibloa{{\tr}_{+}(\islash\, \bslash\, \lslash_1 \, \aslash)}
\def\jaltb{{\tr}_{+}(\jslash\, \aslash\, \lslash_2 \, \bslash)}
\def\ialtb{{\tr}_{+}(\islash\, \aslash\, \lslash_2 \, \bslash)}
\def\bltloa{{\tr}_{+}(\bslash\, \lslash_2\, \lslash_1 \, \aslash)}
\def\jbloa{{\tr}_{+}(\jslash\, \bslash\, \lslash_1 \, \aslash)}
\def\ibPb{{\tr}_{+}(\islash\, \bslash\, \Pslash \, \bslash)}
\def\ijltb{{\tr}_{+}(\islash\, \jslash\, \lslash_2 \, \bslash)}

\def\ijloa{{\tr}_{+}(\islash\, \jslash\,  \lslash_1 \, \aslash)}
\def\ijblt{{\tr}_{+}(\islash\, \jslash\,  \bslash \, \lslash_2)}

\def\jakb{{\tr}_{+}(\jslash\, \aslash\, \kslash \, \bslash)}
\def\iakb{{\tr}_{+}(\islash\, \aslash\, \kslash \, \bslash)}

\def\tofo{{\tr}_{+}(\onslash\, \thslash\, \twslash \, \foslash)}
\def\foto{{\tr}_{+}(\onslash\, \thslash\, \foslash \, \twslash)}
\def\tofi{{\tr}_{+}(\onslash\, \thslash\, \twslash \, \fislash)}
\def\fito{{\tr}_{+}(\onslash\, \thslash\, \fislash \, \twslash)}

\def\lrangle#1#2{\langle #1\,#2\rangle}

\def\Li{{\rm{Li}}}
\def\eps{\epsilon}
\def\epsuv{{\epsilon_{\rm \mbox{\tiny UV}}}}
\let\bm=\bibitem
\let\la=\label

\def\npb#1#2#3{Nucl. Phys. {\bf{B#1}} #3 (#2)}
\def\plb#1#2#3{Phys. Lett. {\bf{#1B}} #3 (#2)}
\def\prl#1#2#3{Phys. Rev. Lett. {\bf{#1}} #3 (#2)}
\def\prd#1#2#3{Phys. Rev. {D \bf{#1}} #3 (#2)}
\def\cmp#1#2#3{Comm. Math. Phys. {\bf{#1}} #3 (#2)}
\def\cqg#1#2#3{Class. Quantum Grav. {\bf{#1}} #3 (#2)}
\def\nppsa#1#2#3{Nucl. Phys. B (Proc. Suppl.) {\bf{#1A}}#3 (#2)}
\def\ap#1#2#3{Ann. of Phys. {\bf{#1}} #3 (#2)}
\def\ijmp#1#2#3{Int. J. Mod. Phys. {\bf{A#1}} #3 (#2)}
\def\rmp#1#2#3{Rev. Mod. Phys. {\bf{#1}} #3 (#2)}
\def\mpla#1#2#3{Mod. Phys. Lett. {\bf A#1} #3 (#2)}
\def\jhep#1#2#3{J. High Energy Phys. {\bf #1} #3 (#2)}
\def\atmp#1#2#3{Adv. Theor. Math. Phys. {\bf #1} #3 (#2)}
\newcommand{\EQ}[1]{\begin{equation} #1 \end{equation}}
\newcommand{\AL}[1]{\begin{subequations}\begin{align} #1 \end{align}\end{subequations}}
\newcommand{\SP}[1]{\begin{equation}\begin{split} #1 \end{split}\end{equation}}
\newcommand{\ALAT}[2]{\begin{subequations}\begin{alignat}{#1} #2 \end{alignat}
                        \end{subequations}}
\def\beqa{\begin{eqnarray}}
\def\eeqa{\end{eqnarray}}
\def\beq{\begin{equation}}
\def\eeq{\end{equation}}
\def\sst{\scriptscriptstyle}
\def\thetabar{\bar\theta}
\def\Tr{{\rm Tr}}
\def\one{\mbox{1 \kern-.59em {\rm l}}}
 \def\Nh{\hat{N}}

\newcommand{\half}{{\textstyle \frac{1}{2}}}

\def\a{\alpha}      \def\da{{\dot\alpha}}
\def\b{\beta}       \def\db{{\dot\beta}}
\def\c{\gamma}  \def\G{\Gamma}  \def\cdt{\dot\gamma}
\def\d{\delta}  \def\D{\Delta}  \def\ddt{\dot\delta}
\def\e{\epsilon}        \def\vare{\varepsilon}
\def\f{\phi}    \def\F{\Phi}    \def\vvf{\f}
\def\h{\eta}
\def\k{\kappa}
\def\l{\lambda} \def\L{\Lambda}
\def\m{\mu} \def\n{\nu}
\def\o{\omega}
\def\p{\pi} \def\P{\Pi}
\def\r{\rho}
\def\s{\sigma}  \def\S{\Sigma}
\def\t{\tau}
\def\th{\theta} \def\Th{\Theta} \def\vth{\vartheta}
\def\X{\Xeta}
\def\z{\zeta}
\def\de{\partial}

\def\cA{{\cal A}} \def\cB{{\cal B}} \def\cC{{\cal C}}
\def\cD{{\cal D}} \def\cE{{\cal E}} \def\cF{{\cal F}}
\def\cG{{\cal G}} \def\cH{{\cal H}} \def\cI{{\cal I}}
\def\cJ{{\cal J}} \def\cK{{\cal K}} \def\cL{{\cal L}}
\def\cM{{\cal M}} \def\cN{{\cal N}} \def\cO{{\cal O}}
\def\cP{{\cal P}} \def\cQ{{\cal Q}} \def\cR{{\cal R}}
\def\cS{{\cal S}} \def\cT{{\cal T}} \def\cU{{\cal U}}
\def\cV{{\cal V}} \def\cW{{\cal W}} \def\cX{{\cal X}}
\def\cY{{\cal Y}} \def\cZ{{\cal Z}}

\def\ua{\underline{\alpha}}
\def\ub{\underline{\phantom{\alpha}}\!\!\!\beta}
\def\uc{\underline{\phantom{\alpha}}\!\!\!\gamma}
\def\um{\underline{\mu}}
\def\ud{\underline\delta}
\def\ue{\underline\epsilon}
\def\una{\underline a}\def\unA{\underline A}
\def\unb{\underline b}\def\unB{\underline B}
\def\unc{\underline c}\def\unC{\underline C}
\def\und{\underline d}\def\unD{\underline D}
\def\une{\underline e}\def\unE{\underline E}
\def\unf{\underline{\phantom{e}}\!\!\!\! f}\def\unF{\underline F}
\def\unm{\underline m}\def\unM{\underline M}
\def\unn{\underline n}\def\unN{\underline N}
\def\unp{\underline{\phantom{a}}\!\!\! p}\def\unP{\underline P}
\def\unq{\underline{\phantom{a}}\!\!\! q}
\def\unQ{\underline{\phantom{A}}\!\!\!\! Q}
\def\unH{\underline{H}}

\def\As {{A \hspace{-6.4pt} \slash}\;}
\def\bs {{b \hspace{-6.4pt} \slash}\;}
\def\Ds {{D \hspace{-6.4pt} \slash}\;}
\def\ds {{\del \hspace{-6.4pt} \slash}\;}
\def\ss {{\s \hspace{-6.4pt} \slash}\;}
\def\ks {{ k \hspace{-6.4pt} \slash}\;}
\def\ps {{p \hspace{-6.4pt} \slash}\;}
\def\pas {{{p_1} \hspace{-6.4pt} \slash}\;}
\def\pbs {{{p_2} \hspace{-6.4pt} \slash}\;}
\def\Ps {{P \hspace{-6.4pt} \slash}\;}
\def\Qs {{Q \hspace{-6.4pt} \slash}\;}

\def\Fh{\hat{F}}
\def\Vh{\hat{V}}
\def\Xh{\hat{X}}
\def\ah{\hat{a}}
\def\xh{\hat{x}}
\def\yh{\hat{y}}
\def\ph{\hat{p}}
\def\xih{\hat{\xi}}
\def\psit{\tilde{\psi}}
\def\Psit{\tilde{\Psi}}
\def\tht{\tilde{\th}}
\def\lt{\tilde{\lambda}}
\def\hl{\hat{\lambda}}
\def\hlt{\hat{\tilde{\lambda}}}
\def\llt{\tilde{l}}
\def\At{\tilde{A}}
\def\Qt{\tilde{Q}}
\def\Rt{\tilde{R}}
\def\Nt{\tilde{N}}

\def\at{\tilde{a}}
\def\st{\tilde{s}}
\def\ft{\tilde{f}}
\def\pt{\tilde{p}}
\def\qt{\tilde{q}}
\def\vt{\tilde{v}}
\def\nt{\tilde{n}}

\def\delb{\bar{\partial}}
\def\bz{\bar{z}}
\def\bD{\bar{D}}
\def\bB{\bar{B}}

\def\bk{{\bf k}}
\def\bl{{\bf l}}
\def\bp{{\bf p}}
\def\bq{{\bf q}}
\def\br{{\bf r}}
\def\bx{{\bf x}}
\def\by{{\bf y}}
\def\bR{{\bf R}}
\def\bV{{\bf V}}

\def\d{\delta}\def\D{\Delta}\def\ddt{\dot\delta}
\def\pa{\partial} \def\del{\partial}
\def\xx{\times}
\def\uno{\mbox{1 \kern-.59em {\rm l}}}
\def\trp{^{\top}}
\def\inv{^{-1}}
\def\dag{{^{\dagger}}}
\def\pr{^{\prime}}
\def\lan{\langle}
\def\ran{\rangle}
\def\rar{\rightarrow}
\def\lar{\leftarrow}
\def\lrar{\leftrightarrow}
\newcommand{\0}{\,\!}      
\def\one{1\!\!1\,\,}
\def\im{\imath}
\def\jm{\jmath}
\newcommand{\tr}{\mbox{tr}}
\newcommand{\slsh}[1]{/ \!\!\!\! #1}
\def\vac{|0\rangle}
\def\lvac{\langle 0|}
\def\hlf{\frac{1}{2}}
\def\ove#1{\frac{1}{#1}}
\def\Box{\square}
\def\ZZ{\mathbb{Z}}
\def\CC#1{({\bf #1})}
\def\bcomment#1{}
\def\bfhat#1{{\bf \hat{#1}}}
\def\VEV#1{\left\langle #1\right\rangle}
\newcommand{\ex}[1]{{\rm e}^{#1}} \def\ii{{\rm i}}
\def\rr{{\rm r}} \def\rs{{\rm s}}\def\rv{{\rm v}}
\def\ri{{\rm i}}\def\rj{{\rm j}}
\newcommand{\lrbrk}[1]{\left(#1\right)}
\newcommand{\sfrac}[2]{{\textstyle\frac{#1}{#2}}}

\def\cS{{\cal S}} \def\cT{{\cal T}} \def\cU{{\cal U}}
\def\cR{{\cal R}}

\font\mybb=msbm10 at 12pt
\def\bb#1{\hbox{\mybb#1}}

\font\myBB=msbm10 at 18pt
\def\BB#1{\hbox{\myBB#1}}

\begin{document}

\begin{flushright}
IPPP/12/24, DCPT/12/48
\end{flushright}

\vspace{10pt}

\begin{center}

 \hspace{-0.8cm}{\Large \bf Uplifting Amplitudes in Special Kinematics}

\vspace{20pt}
\end{center}

\centerline{\mbox {\large Timothy Goddard$^{a}$,  Paul~Heslop$^{a}$ and
Valentin~V.~Khoze$^{b}$}%
{
\renewcommand{\thefootnote}{}  \footnotetext{
{\tt t.d.goddard@durham.ac.uk, paul.heslop@durham.ac.uk, valya.khoze@durham.ac.uk} } } }

\vspace{5pt}

\begin{center}
{\small \em
\begin{itemize}
\item[\ \ \ \ \ \ $^a$]
Department of Mathematical Sciences\\
Durham University,
Durham, DH1 3LE, United Kingdom\\
\item[\ \ \ \ \ \ $^b$]
Institute for Particle Physics Phenomenology,  \\
Department of Physics,
Durham University, \\
Durham, DH1 3LE, United Kingdom

\end{itemize}
}


\vspace{23pt} {\bf Abstract}
\end{center}

\noindent We consider scattering amplitudes in planar $\cN=4$ supersymmetric Yang-Mills theory in
special kinematics where all external four-dimensional momenta are restricted to a (1+1)-dimensional subspace.
The amplitudes are known to satisfy non-trivial factorisation properties arising from multi-collinear limits, which
we further study here. We are able to find a general solution to these multi-collinear limits. This results in a
simple formula which represents an $n$-point superamplitude in terms of a linear combination of functions $S_m$ which
are constrained to vanish in all appropriate multi-collinear limits. These collinear-vanishing building blocks, $S_m$,
are dual-conformally-invariant functions which depend on the reduced $m$-point kinematics with $8\le m\le 4\ell$.
For MHV amplitudes they can be constructed directly using, for example, the approach in Ref.~\cite{HK3loop}.
This procedure provides a universal uplift of lower-point collinearly vanishing building blocks $S_m$
to all higher-point amplitudes. It works
at any loop-level $\ell \ge 1$ and for any MHV or N$^k$MHV
amplitude. We compare this with explicit examples involving $n$-point
MHV amplitudes at 2-loops and $10$-point MHV amplitudes at 3-loops. Tree-level superamplitudes have different properties and are
treated separately from loop-level amplitudes in our approach. To illustrate this we derive an expression for $n$-point
tree-level NMHV amplitudes in special kinematics.

\thispagestyle{empty}

\newpage

\thispagestyle{empty}

  {\small \tableofcontents}

\setcounter{page}{0} \thispagestyle{empty}
\newpage

\def\cS{{\cal S}}
\def \tens{\otimes}

\pagestyle{plain} \textheight 220mm \textwidth 6.0in
\oddsidemargin .10in \evensidemargin .2in \topmargin -.25in
\headheight 12pt \headsep .275in
\footskip 30pt

\section{Introduction}
\setcounter{footnote}{0}
\setlength{\parskip}{15pt}
Scattering amplitudes in
gauge theory (and gravity) are known to have a much simpler underlying structure than implied by their
direct construction in terms of Feynman diagrams. One theory where these simplifications are
particularly striking is the planar ${\cal N}=4$ supersymmetric gauge theory. In fact, it is not
unreasonable to expect that the entire S-matrix in this theory can one day be determined from methods based
on integrability of planar ${\cal N}=4$ SYM.

In this paper we pursue the approach of using arguments based on symmetry considerations rather than
direct perturbative calculations in establishing the structure of scattering amplitudes. In addition to
simplifications occurring from working in the planar limit of maximally supersymmetric Yang-Mills, we
take an additional step of imposing a kinematical restriction on external momenta of scattered states.
This corresponds to confining all external momenta to live in $1+1$ dimensions of the full $3+1$ dimensional
Minkowski space (the loop momenta of course are not restricted). Based on experience with the currently
available results at up to 2-loops, we know that their analytical form simplifies considerably when one
restricts to this special external kinematics. We thus can think of this restriction as a short cut towards
establishing the underlying integrable structure of the amplitude in full kinematics.

The 2d external kinematics was first introduced
in~\cite{am8} in the context of strong coupling where it was interpreted as the boundary of the $AdS_3$ target space
of the dual string theory. The lowest-$n$ non-trivial amplitudes in this kinematics are 8-point amplitudes
and Ref.~\cite{am8} computed them in the strong coupling regime. {
  At both  weak and strong coupling, maximal
helicity violating (MHV) amplitudes are conjectured to be dual to polygonal
light-like Wilson loops~\cite{AM,dks,bht,7authors,dhks6} and using
this duality,
the 8-point MHV amplitude  was computed in 2d kinematics at 2-loops
in \cite{dds-2d}.}
Then an infinite sequence of MHV 2-loop amplitudes was found in~\cite{HK2loop}
where the 2-loop result for the 8-point
MHV amplitude was extended to all
$n$-points\footnote{$n$ must be even in the 2d kinematics.}, using symmetry and collinear limits as well as an additional
assumption about the analytic structure of the $n$-point answer.
This construction was further sharpened in \cite{HK3loop} where 3-loop expressions (with a few unfixed coefficients)
for MHV amplitudes were obtained
at $8$ points and $10$ points. The main motivation of the present paper is to construct a universal uplift of lower-point amplitudes to
arbitrary number of points $n$. In other words, we want to upgrade the construction of 2-loop MHV $n$-point amplitudes
carried out in \cite{HK2loop} to all-loop MHV and non-MHV $n$-amplitudes in terms of the low-$n$-point building blocks.

\medskip

Colour ordered $n$-point N$^k$MHV scattering amplitudes in planar ${\cal N}=4$ SYM theory, $A_{n,k}$,
are the central objects of interest.
Each N$^k$MHV amplitude $A_{n,k}$ is a combination of all possible
physical amplitudes involving
$k+2$ { negative helicity gluons and the rest  positive helicity
gluons, together with amplitudes related to these by supersymmetry}.
These amplitudes depend
on the on-shell momenta $p_i$ of the external particles and on Grassmann variables $\eta^A_i$
 necessary to specify all the particle states of
the super Yang-Mills multiplet (for example, the positive helicity gluon $g_i^{+}$ is characterised by $\eta_i$ to the
power zero and, on the opposite end of the spectrum,
the negative helicity gluon $g_j^{-}$ corresponds to
$\eta^1_j \eta^2_j \eta^3_j \eta^4_j$, while fermions and scalars fill in powers of $\eta$ from one to three).
Each N$^k$MHV amplitude $A_{n,k}$ is of degree $(\eta)^{8+4k}$,
see \cite{Nair,GGK} for more detail.

All $A_{n,k}$ amplitudes are naturally assembled into a single ${\cal N}=4$ superamplitude,
\be
A_n \,=\,\sum_{k=0}^{n-4} \, A_{n,k} \, ,
\ee
and can be read back from it as $(\eta)^{8+4k}$ coefficients in the Taylor
expansion in terms of the Grassmann variables $\eta^A_i$.

It will be convenient for us to factor out from the superamplitude $A_n$ the tree-level
contribution $A_n^{\rm tree}$, as well as the infrared divergences coming from loops,
\be
A_{n} \, =\, A_{n}^{\rm tree} \, M_n^{\sst BDS}\, R_{n}\, .
\label{Rnkdef}
\ee
Here $M_n^{\sst BDS}$ denotes the known\cite{bds} BDS-expression which contains all infrared
divergences of the amplitude, and it is also known to factorise correctly under simple collinear limits,
where two consecutive momenta become collinear.

Thus through \eqref{Rnkdef} the ${\cal N}=4$ scattering amplitudes are determined in terms of $R_n$ which is
known as the reduced function or the remainder function. In our definition $R_n$ itself is a
superfunction, and can be Taylor-expanded in Grassmann $\eta$'s to give the N$^k$MHV remainder
functions, $R_n = \sum_{k=0}^{n-4} \, R_{n,k}$.
Each $R_{n,k}$ is a finite, regularisation-independent
and dual-conformally invariant quantity. { More precisely, for the MHV
case $k=0$, the amplitude/ Wilson loop duality predicts that $R_{n,0}$
is dual conformally invariant and
depends on external momenta 
only through conformal cross-ratios $u$ \cite{dhks5}.
In the general N$^k$MHV case, dual superconformal
invariance~\cite{Drummond:2008vq}, fully present at
tree-level~\cite{Brandhuber:2008pf} but partially broken at loop
level, implies that $R_{n,k}$
depends on the external kinematics (momenta and helicities)
through the cross-ratios as well as through dual superconformal invariants~\cite{Drummond:2008vq} which involve Grassmann variables.
 There is now a conjectured duality between the superamplitude and a super-Wilson
 loop~\cite{Mason:2010yk,CaronHuot:2010ek} as well as with supersymmetric correlation functions~\cite{EKS2,paper1,paper2}, both of
 which  explain the presence of dual superconformal symmetry.
}

It will be advantageous to consider the logarithm of the reduced superamplitude, $\cR_n=\log R_n$.
In perturbation theory it can be expanded in powers of the coupling, and independently of this also
in powers of $\eta$
\begin{equation}
  {\cR}_{n} \,:=\,\log R_{n} \, =\, \sum_{\ell=1}^\infty\, a^\ell\, \cR_{n}^{(\ell)}
  \, =\, \sum_{k=0}^{n-4}\, \sum_{\ell=1}^\infty\, a^\ell\, \cR_{n,k}^{(\ell)}
  \, .
\label{cRnkdef}
\end{equation}
{ Note that  $\cR_{n,k}$ will thus have contributions from
  $R_{n,k}$ but also from products $R_{n,k'}R_{n,k-k'}$.}
We note that in our definition of $R_n$ in Eq.~\eqref{Rnkdef}
we factored out the entire tree-level superamplitude (rather than, for example, only the MHV expression,
as is sometimes done in the literature, see e.g. \cite{ch-he}).
Thus all tree-level contributions are cleanly separated from loops and the expansion on the {\it r.h.s.}
of \eqref{cRnkdef} starts from $\ell=1$ loop.

For MHV contributions, the expansion starts at 2-loops since
$\cR_{n,0}^{(1)}=0$. { In general four-dimensional kinematics,
non-trivial two-loop  contributions start at 6-points, and
$\cR_{6,0}^{(2)}$ was obtained numerically
in~\cite{7authors,dhks6} and later analytic expressions for
$\cR_{6,0}^{(2)}$ were derived in \cite{6an1,6an2,gsv}. The result
for general $n$,  $\cR_{n,0}^{(2)}$,  can be obtained numerically
from the algorithm constructed in \cite{abhkst}. The
symbol~\cite{gsv} of the $n$-point amplitude, $\cR_{n,0}^{(2)}$,
is known~\cite{CaronHuot:2011ky} as is the symbol of the six-point
3-loop amplitude $\cR_{6,0}^{(3)}$~\cite{Dixon:2011pw} and the
six-point two-loop NMHV amplitude
$\cR_{6,1}^{(2)}$~\cite{Dixon:2011nj}. In special  two-dimensional
kinematics, remarkably compact analytic expressions for
$\cR_{n,0}^{(2)}$ were derived in \cite{dds-2d} at $n=8$, and in
\cite{HK2loop} for all $n$. Three-loop analytic expressions for
$\cR_{n,0}^{(3)}$, containing 7 unfixed coefficients, were
obtained in \cite{HK3loop} for $n=8$ in special 2d kinematics and
further generalised to $n=10$.}

The NMHV amplitudes in special kinematics will be addressed in~\cite{wip}.

The purpose of the present paper is to present a universal method for
uplifting lower-point amplitudes in special kinematics, such as $\cR_{n=8,k}^{(\ell)}$ to all higher $n$ for all classes
of N$^k$MHV's. Specifically we want to construct the uplift of the superamplitude $\cR_n$.

We start in general 4d kinematics and  our first main new
result is a full explication of all $k$-preserving and
$k$-decreasing multi-collinear limits. When written in terms of
the full superamplitude and defined in the correct supersymmetric
way, the collinear limits take on a remarkably simple form, Eqs.~\eqref{superRcoll}-\eqref{superRcolllog}
in Section~{\bf \ref{sec:two}}. Using this form for the collinear limits,
we are able to completely and explicitly solve their consequences
for higher point amplitudes restricted to 2d kinematics, firstly at MHV level in
Section~{\bf \ref{sec:four}} and then for any
superamplitude at N${}^k$MHV level in
Section~{\bf \ref{sec:collinear-uplift-n}}. This form of the collinear
uplift is  the unique one with manifest collinear properties.

Tree-level amplitudes manage to possess correct
collinear limits in a non-manifest way by satisfying non-trivial
linear identities. We illustrate this by giving the NMHV tree-level
$n$-point amplitude in Section~{\bf \ref{sec:tree-level-nmhv}}.

\section{Kinematics and collinear limits}
\label{sec:two}

Superamplitudes are functions of bosonic variables (the light-like momenta $p_i$ of external particles) as well as fermionic
variables $\eta^A_i$ \cite{Nair} which take into account the different states in
the super Yang-Mills multiplet which are being scattered.
All $k$-components of the N$^k$MHV amplitudes $A_{n,k}$ arise from the Taylor
expansion of the superamplitude in terms of the Grassmann variables $\eta^A_i$,
as explained in Sec.5 of Ref.~\cite{GGK}.

It is useful to package the external data $\{p^\mu_i,\eta^A_i\}$ in terms of
the region momenta $x_i^{\a \da}$
and their fermionic components $\theta^{\a A}_i$ defined as follows \cite{Drummond:2008vq}
\bea
p_i^{\a \da} \, \equiv\,  \lambda_i^\a \tilde{\lambda}_i^\da &=& x_i^{\a \da} - x_{i+1}^{\a \da} \, , \qquad \a,\da =1,2\ ,
\label{xdef}\\
\lambda_i^\a \eta^A_i &=& \theta^{\a A}_i - \theta^{\a A}_{i+1} \, , \qquad A=1,\ldots, 4\, ,
\label{thetadef}
\eea
where $\lambda_i^\a$ and  $\tilde{\lambda}_i^\da$ are the standard two-component helicity spinors.
The chiral superspace coordinates $X_i=(x_i,\theta^{A}_i)$ define the vertices of the $n$-sided
null polygon contour for the Wilson loop dual to the $n$-point superamplitude \cite{AM,Drummond:2008vq,bht}.

We will also use momentum supertwistors \cite{Hodges,Mason:2009qx} which transform linearly under $SU(2,2|4)$ dual
superconformal transformations. The supertwistors are defined via
\be
\cZ_i\,=\, (Z_i^a \,; \,\chi_i^A) \,=\, (\lambda_i^\a, x_{\da \a\,i}\lambda_i^\a\, ;\, \theta^{A}_{\a\,i}\lambda_i^\a)\, .
\label{stwistordef}
\ee
where $Z^a$ denote the four bosonic, and  $\chi^A$ are the four fermionic components.

\subsection{Multi-collinear limits}
\label{sec:coll}

Here we will describe how the collinear limits where $m+1$ consecutive momenta with $m\ge 1$ become collinear act on the
reduced superamplitude $R_{n}$ and its logarithm $\cR_{n}$.
The full superamplitude, $A_n$, factorises in the $m+1$ collinear limit as follows,
\be
\label{superAcoll}
A_n \, \rightarrow\,  A_{n-m} \times {\rm Split}_{m} \, ,
\ee
where $A_{n-m}$ is the superamplitude with $n-m$ external states, and
the expression ${\rm Split}_{m}$ denotes
the splitting superamplitude. The latter can be expanded, ${\rm Split}_m = \sum_{p=0}^{m} {\rm Split}_{m,p}$,
in terms of helicity-changing (also called $k$-reducing)
splitting functions so that the N$^k$MHV
amplitude goes to
\bea
A_{n,k} \, &&\rightarrow\, A_{n-m,k} \times {\rm Split}_{m,0}\, +\, A_{n-m,k-1} \times {\rm Split}_{m,1}
\,+\, \ldots \nonumber\\
&&=
\sum_{p=0}^k  A_{n-m,k-p} \times {\rm Split}_{m,p}\, .
\label{kAcoll}
\eea

The simplest collinear limit occurs when just 2 consecutive momenta in the colour-ordered amplitude become collinear.
The amplitude $A_n$ factorises
in this limit into the
amplitude with $n-1$ external particles times the splitting function, $A_n \to A_{n-1}\times {\rm Split}_1$.
It is well-known \cite{bds,Korchemsky:2009hm} that the BDS expression together with the tree-level amplitude, fully account
for the splitting amplitude ${\rm Split}_1$. As a result the reduced superamplitude has a particularly simple
form under this minimal collinear limit,  $\cR_n \to \cR_{n-1}$, i.e. there is no splitting function entering the $\cR$-equation.

As the next step, let us consider the triple collinear limit where $m+1=3$ consecutive momenta
become collinear, and furthermore we require that the helicity of the amplitude is conserved.
Such limits are referred to as $k$-preserving collinear limits, they focus on the $p=0$ term on the
{\it r.h.s} of \eqref{kAcoll}. The new feature of the triple collinear limit compared to the simple collinear limit
before, is that the corresponding splitting function is no longer
fully accounted for by the BDS expression $M^{\sst BDS}$.
When interpreted in terms of the reduced amplitude,
the factorisation theorem for the helicity-preserving triple collinear limit
gives
\be\label{triple-collR}
{\rm lim}_{\sst k {\rm \, fixed}} \, R_{n,k} \, \rightarrow \, R_{n-2,k}\times {\rm split}_{2,0} \, =\,
R_{n-2,k}\times R_{6,0} \, ,
\ee
where ${\rm split}_{2,0}$ is the helicity-preserving triple collinear splitting amplitude (or more precisely, the part
thereof which is not
accounted by the BDS expression). Importantly, this splitting amplitude agrees with the 6-point MHV reduced
amplitude $R_{6,0}$ \cite{7authors,abhkst}.\footnote{Equation \eqref{triple-collR} was originally derived
in the MHV case
$R_{n,0}$. But once we know that the $k$-independent splitting amplitude
${\rm split}_{2,0}$ is equal to $R_{6,0}$, this fact can  be used for general $k$ in $k$-preserving collinear limits,
leading to \eqref{triple-collR}.}

Moving on to $k$-preserving multi-collinear limits with
$(m+1)$-collinear momenta. Here we have
\be
\label{log-split}
{\rm lim}_{\sst k {\rm \, fixed}} \, R_{n,k}  \,\rightarrow\, R_{n-m,k}\times R_{m+4,0} \, ,
\ee
where similarly to \eqref{triple-collR} the splitting amplitude becomes
the reduced amplitude $R_{m+4,0}$ itself \cite{HKregular,HK2loop}.

We are now ready to consider the general multi-collinear case, where we no longer
impose any restrictions on preserving the helicity degree $k$ of the amplitude.
Thus we write the analogue of the superamplitude factorisation \eqref{superAcoll}
directly for the reduced superamplitude,
\be
\label{superRcoll}
R_n \, \rightarrow\,  R_{n-m} \times R_{m+4} \, .
\ee
This formula can also be expanded in terms of  N$^{k-p}$MHV components similarly to \eqref{kAcoll}
except that now all the splitting-function contributions are expressed in terms of $R$'s:
\bea
R_{n,k} \, &&\rightarrow\, R_{n-m,k} \times R_{m,0}\, +\, R_{n-m,k-1} \times R_{m,1}
\,+\, \ldots \nonumber\\
&&=
\sum_{p=0}^k  R_{n-m,k-p} \times R_{m,p}\, .
\label{kRcoll}
\eea
The $k$-preserving collinear limit \eqref{log-split} is a special case
of these general relations which corresponds to a single term on the
{\it r.h.s.} of \eqref{kRcoll}.

The proof of this collinear factorisation for $R_n$  in \eqref{superRcoll} uses known universal
collinear factorisation properties of amplitudes, combined with dual
superconformal symmetry of $R_n$. We know that
the superamplitude $A_n$ has universal collinear factorisation
limits~\eqref{superAcoll} and so does $M_{BDS}$ being the exponent of the 1-loop MHV amplitude. Therefore the reduced
amplitude $R_n$ as defined in~\eqref{Rnkdef} must also have universal collinear factorisation
properties. Thus we only need to discover what the corresponding splitting superamplitude
is. To do this let us focus
 on the {\it maximal} multi-collinear limit where $n=m+4$.
In this limit from universal factorisation we have
$R_{m+4}\rightarrow R_4 \times {\rm split}_m \,=\, {\rm split}_m $ since $R_4$ is trivial.
On the other hand,
the same $(m+1)$-collinear limit on $R_{m+4}$ can be achieved via a superconformal
transformation on all $m+4$ points (as we show in Appendix A). Therefore we
have $R_{m+4} \rightarrow R_{m+4}$ in this case. The conclusion is
that the splitting amplitude is ${\rm split}_m=R_{m+4}$ and Eq.~\eqref{superRcoll} follows.

Taking the logarithm we get the linear realisation of multi-collinear limits,
\be
\label{superRcolllog}
\cR_n \, \rightarrow\,  \cR_{n-m} \,+\, \cR_{m+4} \, .
\ee
Equations \eqref{superRcolllog} or \eqref{superRcoll} constitute our main result as far as general collinear limits
are concerned, and they will play a key role in in constructing the uplift to general $n$
of the amplitude in the 2d external kinematics as will be introduced below.

\subsection{Two-dimensional kinematics}

In this
subsection we give details and conventions for the special kinematics, first introduced in \cite{am8},
where the external momenta $p_i$ lie entirely in
$1+1$ dimensions.

\begin{figure}
  \centering
\includegraphics[height=8cm]{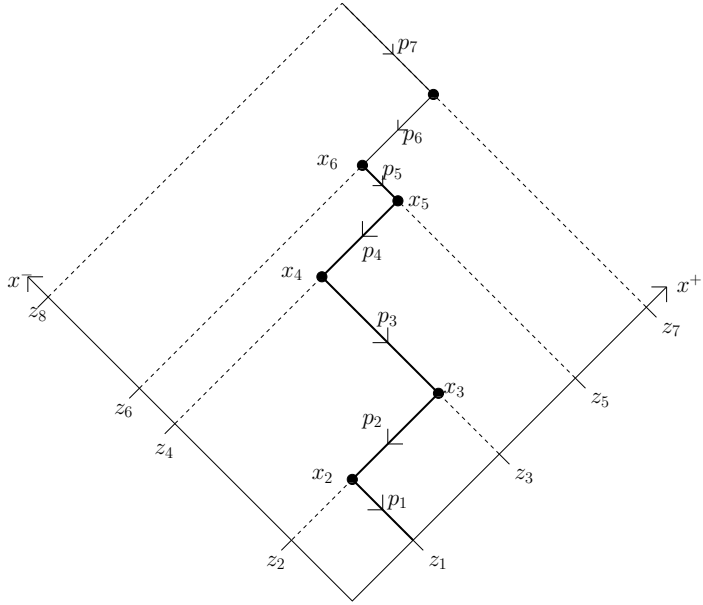}
  \caption{Figure illustrating the zig-zag Wilson loop contour in 2d
    kinematics. Vertices $x_i$  are defined in terms of light-cone
    co-ordinates. In 2d the contour can also be specified by giving every
    other vertex $x_2, x_4, x_6, \dots$.  }
  \label{Fig:zig-zag}
\end{figure}

The corresponding contour of the dual Wilson loop has a zig-zag shape which is shown on the
lightcone plane in Fig.~\ref{Fig:zig-zag}.
The region momenta $x_i$ (vertices of the corresponding Wilson loop contour) have the
following form in light-cone coordinates $(x_+,x_-)$:
\begin{align}\label{eq:6}
  x_{i}=\left\{
  \begin{array}{ll}
(z_{i-1},z_{i})\ , \qquad &i\ \rm{even}\\
       (z_{i},z_{i-1})\ , \qquad &i\ \rm{odd}
     \end{array}\right.
\end{align}
Only an even number of vertices is possible in this 2d kinematics,
and we continue denoting it as $n$ (rather than $2n$ as sometimes done in the literature).
In our notation $z_i$ components with odd values of $i$ lie along the $x^+$ axis, and the
even $z_i$'s are along the $x^-$ axis, as one can see instantly from Fig.~\ref{Fig:zig-zag}.
We will
frequently refer to them as `odd' and `even' coordinates.
All the (bosonic) functions we consider can be written in terms of Lorentz invariant intervals $z_{ij}:=z_i-z_j$
where both $i,j$ are either even or odd.
In this notation we must remember that the even and odd
coordinates are independent of each other.

It is instructive to view the 2d kinematics from the point of view of
momentum twistors \eqref{stwistordef}. In 2d the bosonic twistors
$Z_i^a= (\lambda_i^\a, x_{\da \a\,i}\lambda_i^\a)$ reduce as follows.
For all even values of $i$ we have
\be
\label{even1def}
p_i^{\a \da} \,=\,
\left(\begin{matrix}
 0 & 0 \\ 0 & p_i^-
\end{matrix} \right)
\,=\,  \lambda_i^\a \tilde{\lambda}_i^\da \quad \Rightarrow \quad
\lambda_i^\a = \left(\begin{matrix}
 0  \\ 1
\end{matrix} \right) \,\, , \,\,
\tilde{\lambda}_i^\da = \left(\begin{matrix}
 0  \\ p_i^-
\end{matrix} \right) \, ,
\ee
and
\be
\label{even2def}
x_{\da \a\,i}\lambda_i^\a \,=\,
\left(\begin{matrix}
 x_i^+ & 0 \\ 0 & x_i^-
\end{matrix} \right)\left(\begin{matrix}
 0  \\ 1
\end{matrix} \right)\, =\, \left(\begin{matrix}
 0  \\ z_i
\end{matrix} \right)
 \, .
\ee
For odd values of $i$ the story is similar, and as a result, momentum twistors in
2d have a checkered pattern:
\begin{align}
\label{2dtwist}
  Z_{i}=\left\{
  \begin{array}{ll}
(Z_i^{1},0,Z_i^{3},0)=(1,0,z_i,0) &\qquad i\ \rm{odd}\\
(0,Z_i^{2},0,Z_i^{4})=(0,1,0,z_i) &\qquad i\ \rm{even}\ ,
\end{array}
\right.
\end{align}
which is a manifestation of $SU(2,2) \to SL(2)_+ \times SL(2)_-$ in 2d.

In 2d kinematics it is then natural to define an $SL(2)_{\pm}$-invariant two-bracket
of twistors,
\begin{align}
\label{2brack}
  \vev{ij}:= \left\{
  \begin{array}{lll}
 Z_i^{3}Z_j^{1}-Z_i^{1}Z_j^{3} &\qquad i\ {\rm and}\ j \ \rm{odd}\\
Z_i^{4}Z_j^{2}-Z_i^{4}Z_j^{2} &\qquad i\ {\rm and}\ j \ \rm{even}\\
0 &\qquad \rm{otherwise}
\ .
\end{array}
\right.
\end{align}
From \eqref{2brack} and the {\it r.h.s.} of \eqref{2dtwist} we have that $\vev{ij}=z_{ij}$
and the Lorentz-invariant intervals $z_{ij}$ have the standard two-bracket interpretation $\vev{ij}$
(but in terms of reduced 2d twistors rather than helicity spinors).

Furthermore, the standard $SL(2,2)$-invariant twistor 4-bracket contraction,
\begin{align}\label{eq:3}
  \vev{ijkl}:=\epsilon_{abcd} Z_i^a Z_j^b Z_k^c Z_l^d\ ,
\end{align}
reduces in 2d to a product of two-brackets if
there are two even and two odd indices, or vanishes otherwise:
e.g.  $\vev{1234}=\vev{13}\vev{24}$.
The main point here is that lightcone coordinates are interchangeable with twistors in 2d
and only two-brackets of bosonic twistors (of the same parity) can appear.

For superamplitudes in 2d, it is natural to consider a supersymmetric reduction,
$SU(2,2|4) \to SL(2|2)_+ \times SL(2|2)_-$, under which
momentum supertwistors \eqref{stwistordef} become \cite{ch-he}
\begin{align}
\label{2dstwist}
  \cZ_i\,=\, (Z_i^a \,; \,\chi_i^A) =\left\{
  \begin{array}{ll}
(Z_i^{1},0,Z_i^{3},0;\chi_i^{1},0,\chi_i^{3},0) &\qquad i\ \rm{odd}\\
(0,Z_i^{2},0,Z_i^{4};0,\chi_i^{2},0,\chi_i^{4}) &\qquad i\ \rm{even}\ ,
\end{array}
\right.
\end{align}
{and we will indeed mostly consider this additional reduction in
  fermionic co-ordinates also. On the other hand one should beware that
while we may still compute meaningful forms for either $R_{n,k}$ or $\mathcal{R}_{n,k}$ the MHV-prefactor to this from \eqref{Rnkdef} contains
\begin{equation}
\delta^{(8)} \left( \sum_{i=1}^{n} \lambda_i \eta_{i} \right)
\end{equation}
which under this $\mathrm{SU}(4)$ splitting necessarily goes to
zero. For $R_{n,k}$ with $k=0,1$ this reduction in the fermionic superspace co-ordinates  does
not have a great effect. All results obtained can straightforwardly be
uplifted to the case with full fermionic
dependence.\footnote{I.e. non-vanishing $(\chi^1, \chi^2,
  \chi^3,\chi^4)$ in both equations~\eqref{2dstwist}.} Beyond NMHV
this restriction however does mean a loss of information.}

The remainder function $\cR_{n}$ is dual conformally invariant~\cite{dks,dhks5}
and as such its lowest bosonic component, $\cR_{n,0}$, can be written as a function of cross-ratios.
The non-MHV components, $\cR_{n,k>0}$, also depend on superconformal invariants involving Grassmann variables.
We will first concentrate on the purely bosonic case.

We define the most general cross-ratios in special 2d kinematics as
\begin{align}\label{uijkl-def}
  u_{ij;kl}= {\vev{il}\vev{jk} \over \vev{ik}\vev{jl}}\, .
\end{align}
This equation is meaningful only for $i,j,k,l$ having the same parity, in other words,
all the cross-ratios fall into two separate classes with all indices being even or with all
indices odd. Cross-ratios with indices of mixed parity (even and odd) don't exist.

The general cross-ratios in 2d kinematics have to satisfy an additional constraint,
\be \label{constr-def}
u_{ij;kl} \,=\, 1 - u_{il;kj}\, .
\ee

The set of general $u_{ij;kl}$'s can be reduced to a smaller set
of cross-ratios with only two indices,
\begin{equation}
  \label{eq:uij}
u_{ij} =
\left\{
  \begin{array}{ll}{    \vev{i-1,j+1}\vev{i+1,j-1}
\over
    \vev{i-1,j-1}\vev{i+1,j+1}}=u_{i-1,i+1;j-1,j+1} \qquad \qquad & |i-j|\geq 3, \ i,j\text{ of the same parity}\\
    \nonumber\\
1 \qquad & i,j\ \text{of opposite parity}\ .
  \end{array}
\right.
\end{equation}
Indeed we have
\be\label{udecomp}
  u_{ij;kl}=\prod_{I=i+1}^{j-1} \prod_{K=k+1}^{l-1} u_{IK}\, .
\ee
This reduced set of $u_{ij}$ cross-ratios also has a 4d interpretation,
$u_{ij}=  {x_{i, j+1}^2 x_{i+1,j}^2 \over x_{i,j}^2 x_{i+1, j+1}^2}$.

For the two lowest-$n$ cases, the octagon and the decagon, all non-trivial 2-component cross-ratios
are of the form $u_{i,i+4}$, with $i=1,\ldots,4$ for the octagon, and $i=1,\ldots,10$ for the decagon.
The cross-ratios $u_{ij}$ are still not all independent because of
equations \eqref{constr-def}, leaving $n-6$ (i.e. 2 for the octagon and 4 for the decagon) independent solutions.
For the octagon \eqref{constr-def} amounts to
\be \label{n8const}
n=8\ : \qquad 1- u_{i,i+4} = u_{i+2,i+6} \, , \qquad i=1,2 \, ,
\ee
and for the decagon,
\be \label{n10const}
n=10\ : \qquad 1- u_{i,i+4} = u_{i+2,i+6} \,u_{i-2,i+2} \,  .
\ee
To simplify notation at low $n$, it is sometimes convenient to use
\be
u_i \,:= u_{i,i+4}\, .
\ee
While at $n=8$ and $n=10$ these are the only cross-ratios, this is no longer true at $n \ge 12$ where
$u_{ij}$ cross-ratios appear with $j-i \ge 6$.
More details on the cross-ratios in the special kinematics can be found
in~\cite{HK2loop}.

\subsection{Collinear limits in the 2d kinematics}

\begin{figure}
  \centering
\includegraphics[height=5cm]{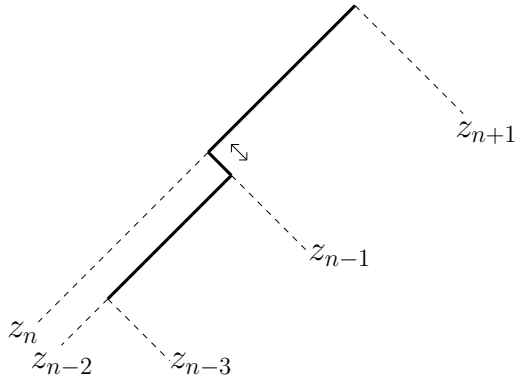}
\caption{Figure illustrating the triple/soft collinear limit $z_n \to z_{n-2}$ while the variable $z_{n-1}$ remains
free. }
  \label{Fig:collinear}
\end{figure}

From the zig-zag kinematics it is clear that the lowest collinear limit one can apply and remain within the
$(1+1)$-dimensional kinematics is the triple collinear limit where three consecutive edges collapse into one.
More precisely, this should be thought of as a collinear-soft-collinear limit, where three edges
with momenta $p_{n-2}$, $p_{n-1}$ and $p_{n}$, collapse into a single edge $p_n$. In practice,
the middle momentum becomes soft, $p_{n-1}\to 0.$ In terms of twistors, or the lightcone components $z_i$'s,
we see that $z_n \to z_{n-2}$ while the variable $z_{n-1}$ remains
free, see figure~\ref{Fig:collinear}.

We now recall that in the 2d  kinematics there are no non-trivial cross-ratios at 6-points (lowest non-trivial case being
$\cR_{8,0}$) and the 6-point reduced amplitude $\cR_{6}$ is a
(coupling dependent) constant
multiplied by the tree-level amplitude, which can be reabsorbed into $\cR_{n}$ which
we now call $\tilde{\cR}_{n}$ \cite{HK3loop},
\be
\label{tildeR-def}
\tilde{\cR}_{n} \,=\, \cR_n - \frac{n-4}{2}\,  \cR_6 \, ,
\ee
so that at the level of superamplitudes we have
\be\label{triple-collctR2d}
 \tilde{\cR}_{n} \, \rightarrow \, \tilde{\cR}_{n-2} \, .
\ee
Also, for $\cR_n = \log R_n$ expressions at different order in the loop expansion do not mix,
\bea
\tilde{\cR}_{n}^{(\ell)}  &\rightarrow& \tilde{\cR}_{n-2}^{(\ell)} \, , \\
\tilde{\cR}_{n}^{(\ell)}  &\rightarrow& \tilde{\cR}_{n-m}^{(\ell)} + \tilde{\cR}_{m+4}^{(\ell)}\, , \quad {\rm for} \ m\ge 4 \, ,
\eea
and thus $\tilde{\cR}_n$ is the natural object to use for collinear uplifts of amplitudes to higher number of points.

\section{$n$-point MHV amplitudes: Part 1}

\label{sec:revi-mhv-ampl}

At 2-loop level, MHV amplitudes in 2-dimensional external kinematics are known for any number of external particles
and, remarkably, the result for $n$-point amplitudes depends only on the four-logarithms structure appearing at 8-points.
If we denote the 8-point 2-loop structure $S_8^{(2)}$, the general $n$-point remainder function $\tilde R_n^{(2)}$ at 2 loops
emerges as a linear combination of $S_8^{(2)}$s and nothing else. Can this be the case at 3 loops?

The MHV amplitudes at 3-loops were constructed for $n=8$ and 10 points. In the latter case the 10-point expression contained
the 8-point structures $S_8^{(3)}$, as well as new non-trivial contributions which did not appear at 8 points. Beyond the 10-point case,
the structure of $n$-point 3-loop amplitude was until now an open problem.

The goal of this and the following sections is to find a decomposition of the general $n$-point amplitude $\tilde R_n^{(\ell)}$
in terms of the lower-point building blocks, $S_m$, with trivial multi-collinear limits.
This formula will be given in the following section.
Below we will briefly recall the known results about 2-loop and 3-loop amplitudes and then proceed to recast them in the form
which is more appropriate to the $\sum S_m$ generalisation.

\subsection{n-points at 2 loops}

In~\cite{HK2loop} the $n$-point 2 loop MHV amplitude was obtained in
2d kinematics.
The result can be written entirely in terms of
logarithms of the simple cross-ratios $u_{ij}$ and takes the form\footnote{For MHV amplitudes
we will suppress the $k=0$ subscript to simplify notation.}
\begin{align}
  \label{Rn2-result}
  R^{(2)}_n\,:= \,R^{(2)}_{n,0} &= -{1 \over 2} \Big( \sum_{\cS} \log ( u_{i_1 i_5}) \log
  ( u_{i_2 i_6}) \log ( u_{i_3 i_7}) \log ( u_{i_4 i_8}) \Big) -
  {\pi^4\over 72} (n-4)\ ,
\end{align}
where the sum runs over the set
\begin{align}
  \cS &= \Big\{ i_1, \dots i_8: 1\leq i_1<i_2< \dots < i_8 \leq
  n,\qquad i_k -i_{k-1} = \mathrm{odd} \Big\}\ .
\end{align}
The constant term on the {\it r.h.s.} of \eqref{Rn2-result} arises from $R^{(2)}_{6,0}$ and can be removed
by going to $\tilde{R}_{6}$ as in \eqref{tildeR-def}. Also, for ease of dealing with collinear limits
it is more appropriate to work with the logarithm of the amplitude
$\cR_n$, though for MHV expressions the first differences would
appear starting from 4-loops.
We thus will use
\be
\label{rn}
 \tilde \cR^{(2)}_{n} = -{1 \over 2} \Big( \sum_{\cS} \log ( u_{i_1 i_5}) \log
  ( u_{i_2 i_6}) \log ( u_{i_3 i_7}) \log ( u_{i_4 i_8}) \Big) \, .
\ee

The 2-loop $n$-point result was found in \cite{HK2loop} by examining the consequences of
collinear limits described above, starting with the 8-point amplitude computed (via the
Wilson loop/amplitude duality) in~\cite{dds-2d} and first using an additional assumption that only
logarithms of simple cross-ratios $u_{ij}$ can appear at two loops. The general 2-loop
analytic formula (and thus the logs-only asumption) was verified \cite{HK2loop} numerically using the code
developed in \cite{abhkst}.

At higher loops, the logs-only structure no longer holds \cite{HK3loop}, and furthermore, the
problem of how to obtain the all-$n$ amplitudes starting from
low-$n$ expressions was previously an open problem.

\subsection{8- and 10-point MHV amplitudes at 3 loops}

The 8-point MHV amplitude $\cR_{8}^{(3)}$ at 3 loops was determined in Ref.~\cite{HK3loop}.
This derivation was based on the fundamental assumption that the amplitude has a symbol whose entries are
cross-ratios\footnote{At 1 and 2-loops this amounts to logarithms only in the amplitude \cite{HK3loop}
and starting from 3-loops the reconstructed functions involve also polylogarithms $Li_n(u_{ij})$.}.
The reader is referred to Appendix B for more detail on this construction.

At 8-points in special kinematics, there are four non-trivial
cross-ratios, $u_1:=u_{15},\  u_2:=u_{26},\ u_3:=1-u_1,\ u_4:=1-u_2$.
Insisting that the 8-point function be cyclically (and parity)
symmetric, and that it vanishes in the collinear limit $z_8 \rightarrow
z_{6}$, i.e. $u_1\rightarrow 0\,, u_3\rightarrow 1$ with $u_2,u_4$
unconstrained\footnote{Note that we need not consider cyclically
  equivalent collinear limits $z_i \rightarrow z_{i-2}$, since they
  will follow automatically from cyclic symmetry.}  leads to a 3-loop
amplitude of the form:
\begin{align}\label{3loopamp}
  \tilde \cR_8^{(3)}=\sum_{\sigma,\tau} a_{\sigma\tau} f_\sigma(u_1)
  f_\tau(u_2)
\end{align}
where $a_{\sigma\tau}=a_{\tau\sigma}$ are some rational coefficients,
and the sum is over the set of functions $f_\sigma$ with the following
properties:
\begin{align}
  f_\sigma(u)&=f_\sigma(1-u)\nonumber\\
  f_\sigma(0)&=0\nonumber\\
  f_\sigma(u) & \text{ is a (generalised) polylogarithm.}\label{eq:9}
\end{align}
Furthermore the total polylog weight (or ``degree of
transcendentality'') of $\tilde \cR_8^{(3)}$ must be six due to the
uniform transcendentality property of perturbative amplitudes in $\cN$=4
SYM.

In~\cite{HK3loop} all possible functions $f_\sigma$ were listed (see also Appendix B where these functions
are called $f^+_\sigma$). There
is a unique weight-two function $f(u)=\log u \log (1-u)$, 3 weight-three functions, and
7 weight-four functions, leading to a total of 13
a priori unfixed coefficients $a_{\sigma\tau}$. Further constraints arise from the OPE analysis
of~\cite{OPE2} which fix 6 of these, leaving 7 unfixed
coefficients \cite{HK3loop}.\footnote{ It is tempting to assume a further
  simplification of the structure, namely that the $f_\sigma$ are of weight 3
  only. This would be consistent with all currently known facts and would leave
  just 3 unfixed coefficients. We will not be making this assumption in the present paper.}

The form of the 8-point amplitude~(\ref{3loopamp}) generalises
straightforwardly beyond 3 loops by simply allowing the functions
$f_\sigma$ to have more general weight, such that the total weight is
$2 \ell$.
\begin{align}\label{lloopamp}
  \tilde \cR_8^{(\ell)}=\sum_{\sigma,\tau} a_{\sigma\tau}
  f^{(\ell)}_\sigma(u_1) f^{(\ell)}_\tau(u_2)
\end{align}
It is also valid at two loops where there is only one allowed function
(up to a multiplicative constant), $f^{(2)}(u)=\log u \log(1-u)$, and we
reproduce the original two-loop result at 8-points found in~\cite{dds-2d}
\begin{align}\label{r8}
  \tilde R_{8}^{(2)}= -\frac{1}{2} \log(u_1) \log(u_2) \log(u_3)
  \log(u_4) \ .
\end{align}

Let us first
consider the uplift of the 8-point amplitude to 10-points following~\cite{HK3loop}.  The idea is to write down
all 10-point functions which
reduce to the 8-point amplitude in the triple collinear limit, plus an additional
contribution which is required to vanish in all such limits.
This lead to\footnote{The original derivation in~\cite{HK3loop} was performed at 3-loops,
  but the resulting expression (in terms of the functions $f^{(\ell)}$) remains valid at $\ell$ loops.}
\bea
  \tilde \cR^{(\ell)}_{10}=&& {1 \over 2} \sum_{\sigma,\tau}
  a_{\sigma\tau} \,\Big(
  f^{(\ell)}_\sigma(u_1)f^{(\ell)}_\tau(u_2)-f^{(\ell)}_\sigma(u_1)f^{(\ell)}_\tau(u_4)+{1\over
    2}f^{(\ell)}_\sigma(u_1)f^{(\ell)}_\tau(u_6)\Big) + \mbox{cyclic
  } \nonumber \\
  &&+ V_{10}^{(\ell)}\ .
  \label{eq:14}
\eea
The last term, $V_{10}^{(\ell)}$, denotes
a generic 10-point function which vanishes in all
triple collinear limits. It is reproduced in Appendix B at 3-loop level from Ref.~\cite{HK3loop}.

The construction of the non-vanishing contribution under triple collinear limits (everything apart from
$V_{10}$) was specific to the case at hand where the 10-point amplitude reduces to the 8-point amplitude.
If we want to uplift \eqref{eq:14} to 12 and higher points, we need to
come up with a more general procedure.

Note that the general 10-point expression has a more complicated structure than
the result at two loops:
\be\label{r10}
  \tilde \cR^{(2)}_{10} = -\frac{1}{2} \Big(\log(u_1) \log(u_2) \log(u_3)
  \log(u_4) \,+\, {\rm cyclic}\Big) \ .
\ee
The reason for this simplification is the simple form of the 2-loop function
$f^{(2)}(u)=\log u \log(1-u)$. Using this together with the fact that
at 10 points $1-u_1= u_7 u_3$ and cyclic, one can check that all the
minus signs in~(\ref{eq:14}) (for $\ell=2$) disappear and the result
reduces to~(\ref{r10}).

\subsection{$8-$ and $10-$points recast}
\label{sec:recast}

We can completely and explicitly solve the constraints coming from
collinear limits at 3-loops in terms of three  structures, related to
the 8-,10- and 12-point amplitudes and more generally at $\ell$-loops
in terms of the $m$-point functions $S_m$ with $m\le 4\ell$. But first, to motivate the general formula, we will recast
the 8- and 10-point amplitudes in a form more suitable for
generalisation, and in the process introduce the new concepts we will need.
Also for pedagogical reasons, here in this subsection, we will follow a simplified
approach for relating the $8$-point functions $S_8$ directly with the amplitudes
$\cR_8$. In the following section we will restore to the general case.

Our first step is to recast the problem back as a function of $z$'s,
that is,
\begin{align}
  \cR_8(u_1,u_2)=\cR_8(z_1,z_2,z_3,z_4,z_5,z_6,z_7,z_8)\ .
\end{align}
Now, on attempting to lift this to higher points, we notice that
in the higher point functions the $z$'s always appear in consecutive
pairs, but with the odd element of the pair always appearing before
the even element. This is exactly what happens in the definition of
 $x_i$ in terms of $z_i$ in~\eqref{xdef}. It suggests that we
further think of the amplitude as a function of position coordinates
$x$'s instead of $z$'s so that:
\begin{align}
 2 S_8(x_i, x_j, x_k, x_l):=  \tilde
  \cR_8(x_i^+,x_i^-,x_j^+,x_j^-,x_k^+,x_k^-,x_l^+,x_l^-)\ ,
\end{align}
which implies,
\begin{align}
 2 S_8(x_2, x_4, x_6, x_8)=  \tilde \cR(z_1,z_2,z_3,z_4,z_5,z_6,z_7, z_8)\ .
\end{align}
Here we have introduced the function $S_8$ of $x$-variables which for
the moment we identify with $\cR_8$.\footnote{In the following section we will in fact generalise
this definition by including an additional contribution to $S_8$ which is distinct from the 8-point amplitude $\tilde \cR_8$.}
Thus via $S_8$ we are defining the Wilson loop zig-zag contour (see
Fig.~\ref{Fig:zig-zag}) by specifying every second vertex.  Let us examine the symmetries of the function
$S_8(x_2, x_4, x_6, x_8)$.  The symmetries of the Wilson loop, $\tilde \cR(z_1, \dots z_8)$
namely cyclic symmetry $C_n$, under which each $z_i \to z_{i+1}$, and parity of the 8-point Wilson loop,
$\tilde \cR(z_1, \dots z_8) \to \tilde \cR(z_8, \dots z_1)$,
give the following
\begin{align}
  S_8(x_2, x_4, x_6, x_8)&=  S_8(x_4, x_6, x_8, x_2)\,=\, S_8(x_6, x_8, x_2, x_4)\,=\, S_8(x_8, x_2, x_4,x_6)\notag\\
  S_8(x_2, x_4, x_6, x_8)&=  S_8(x_{1}^f, x_{3}^f, x_{5}^f, x_{7}^f)\notag\\
  S_8(x_2, x_4, x_6, x_8)&= S_8(x_{8}^f, x_{6}^f, x_{4}^f, x_{2}^f)\
  .\label{eq:5}
\end{align}
Here the first equation follows from cyclicity in $z$ applied twice, i.e. $z_i \to z_{i+2}$. The second equation
is a consequence of $z_i \to z_{i+1}$.
In the last two equations we have defined the flipped $x$ position
\begin{align}
  x=(x^+,x^-) \quad \Rightarrow \quad x^f=(x^-,x^+)\ .
\end{align}
This is necessary in order to properly define the cyclic symmetry in
terms of the $x$-variables.

Interestingly, for the 8-point amplitude in the form ~(\ref{lloopamp})
there exists an additional discrete symmetry -- the flip symmetry -- where each $x$-argument
of $S_8$ becomes flipped,
\be\label{flip-def}
 S_8(x_i, x_j, x_k, x_l) \,=\,
S_8(x^f_i, x^f_j, x^f_k, x^f_l)
\ee
despite the fact that it is not an expected symmetry of the
Wilson loop contour.
To identify this symmetry consider $S_8$ of even arguments,
\be \label{S8oddF}
2S_8(x_2,x_4,x_6,x_8)\,=\,
\tilde \cR(z_1,z_2,z_3,z_4,z_5,z_6,z_7, z_8)\,=\,
\sum_{\sigma,\tau} a_{\sigma\tau} f_\sigma(u_1) f_\tau(u_2)
\ee
and compare it with $S_8$ written in terms of the same variables being flipped,
\be \label{S8oddUF}
2 S_8(x^f_2,x^f_4,x^f_6,x^f_8)\,=\,\tilde \cR_8(z_2,z_1,z_4,z_3,z_6,z_5,z_8,z_7)\,=\,
\sum_{\sigma,\tau} a_{\sigma\tau} f_\sigma(u_2) f_\tau(u_1)
\ee
To understand the right hand side, note that
cross-ratios $u_1=u_{15}$ and $u_2=u_{26}$ by definition \eqref{eq:uij} depend only on even or on odd $z$-variables
respectively,
\begin{equation}
  \label{eq:u1-u2}
u_{1} \,=\,
 \frac{\vev{86}\vev{24}}{\vev{84}\vev{26}}\,=\, \frac{z_{86}z_{24}}{z_{84}z_{26}}\, , \qquad
 u_{2} \,=\,
 \frac{\vev{17}\vev{35}}{\vev{15}\vev{37}}\,=\,
 \frac{z_{17}z_{35}}{z_{15}z_{37}} \, ,
\end{equation}
hence the distribution of $z_i$'s inside $\tilde \cR_8$ in Eqs.~(\ref{S8oddF},\ref{S8oddUF})
implies that these two equations are related by $u_1 \leftrightarrow u_2$.
The symmetry $a_{\sigma \tau}=a_{\tau \sigma}$ with the summation over all functions $f_\sigma$ and $f_\tau$ implies that
the resulting expressions are
symmetric under the interchange $u_1 \leftrightarrow u_2$ and equation \eqref{flip-def} follows.

\medskip

Note also that $S_8$ satisfies the following properties under
the collinear limit $z_8\rightarrow z_6$ (ie $x_8\rightarrow x_7$ as can be seen immediately from Fig.~\ref{Fig:zig-zag}):
\begin{align}
  \lim_{x_8 \rightarrow x_7} S_8(x_2,x_4,x_6,x_8)&=0\
  ,\label{Sids}
\end{align}
or more generally/ geometrically
\begin{align}\label{eq:11}
  S_8(x_i,x_j,x_k,x_l)=0 \qquad \text{if} \qquad x_k, x_l \text{
    are light-like separated}\ .
\end{align}
Having defined the object $S_8$ we now re-examine the $\ell$-loop
10-point amplitude~(\ref{eq:14}),
\begin{align}
   \tilde \cR^{(\ell)}_{10}=\,& {1 \over 4} \sum_{\sigma,\tau}
  a_{\sigma\tau} \,
  f^{(\ell)}_\sigma(u_1)\Big(f^{(\ell)}_\tau(u_2)-f^{(\ell)}_\tau(u_4)+f^{(\ell)}_\tau(u_6)-f^{(\ell)}_\tau(u_8)+f^{(\ell)}_\tau(u_{10})\Big)\notag\\
  &+ \mbox{cyclic
  } + V_{10}^{(\ell)}\ \label{eq:18}
\end{align}
This can be rewritten in terms of $S_8$ in a
suggestive way which will allow generalisation to high $n$-points as
\begin{align}
  \tilde \cR^{(\ell)}_{10}(z_{1},z_{2}, \dots ,z_{8}) &= \sum_{1\leq
    i_1\lhd i_2\lhd i_3\lhd i_4\leq 10} S^{(\ell)}_{8}(x_{i_1},x_{i_2},
  x_{i_3},x_{i_4})(-1)^{i_1+\dots
    i_{4}}\ + V_{10}\ .\label{eq:16}
\end{align}
where, as before $V_{10}$ is an additional collinear vanishing
contribution. The summation convention in this formula will be
explained below~(\ref{eq:15-l}), it basically states that each $i_k > i_{k-1}+1$.

The alternating sign in the sum in this formula combined with the property~(\ref{eq:11}) of $S_8$
are enough to show that this has the right behaviour under collinear
limits and we will see this explicitly below, but more interestingly these observations
lead to  immediate generalisation to higher points and arbitrary loop
order.

\section{$n$-point MHV amplitudes: Part 2}
\label{sec:four}
\subsection{The general formula for the $n$-point collinear uplift }

\label{sec:general-formula-n}

We claim that  the $n$-point MHV amplitude for $\ell \geq 1$, at any loop order is given by
\begin{align}\label{eq:15-l}
  \tilde \cR^{(\ell)}_{n}(z_{1},z_{2}, \dots ,z_{n}) &= \sum_{1\leq
    i_1\lhd  i_2\lhd  i_3\lhd  i_4\leq n} S^{(\ell)}_{8}(x_{i_1},x_{i_2},
  x_{i_3},x_{i_4})(-1)^{i_1+\dots
    i_{4}}\ +\nonumber\\
  &+ \sum_{{1\leq i_1\lhd \dots\lhd i_5\leq n}} S^{(\ell)}_{10}(x_{i_1},x_{i_2},
  \dots
  ,x_{i_{5}})(-1)^{i_1+\dots i_{5}}\ +\nonumber\\
  &+ \sum_{{1\leq i_1\lhd \dots\lhd i_6\leq n}} S^{(\ell)}_{12}(x_{i_1},x_{i_2},
  \dots
  ,x_{i_{6}})(-1)^{i_1+\dots i_{6}}\ +\nonumber\\
  &+ \dots+\nonumber\\
  &+ \sum_{{1\leq i_1 \lhd \dots\lhd i_{2\ell} \leq n}}
  S^{(\ell)}_{4\ell}(x_{i_1},x_{i_2}, \dots ,x_{i_{2\ell}})(-1)^{i_1+\dots
    i_{2\ell}}\ .
\end{align}
Here in order to simplify the notation we have defined the symbol
$\lhd$ as follows \newline
$i \lhd j \Leftrightarrow  i< j-1$. This operation removes terms in
the sum with consecutive $x$'s eg $S_m(\dots, x_i, x_{i+1}, \dots)$.

This is a deceptively simple formula. The full $n$-point
amplitude for arbitrary $n$, and arbitrary loop order is given
explicitly in terms of just $(2\ell-3)$ $m$-point functions,
$S_m,\ m=8,10,12,\dots 4\ell$.
Let us start with the minimal case of $n=8$ external particles. Equation \eqref{eq:15-l}
then implies,
\be \label{R82S8}
\tilde \cR_{8}(z_{1},z_{2}, \dots ,z_{8})\,=\,
S_8(x_2,x_4,x_6,x_8) \,+\, S_8(x_1,x_3,x_5,x_7)\, .
\ee
A simple possibility is that the two terms are in fact the same,
$S_8(x_2,x_4,x_6,x_8) = S_8(x_1,x_3,x_5,x_7)=\frac{1}{2} \cR_{8}$, and this is what we
assumed previously in Section {\bf \ref{sec:recast}}.

There is, however, a more general solution to this
equation where $S_8(x_2,x_4,x_6,x_8) \neq S_8(x_1,x_3,x_5,x_7)$. To examine it, we rewrite
\eqref{R82S8} in terms of $z$-variables,
\be \label{R82S8-2}
\tilde \cR_{8}(z_{1},z_{2}, \dots ,z_{7},z_{8})\,=\,
S_8(z_{1},z_{2}, \dots , z_{7},z_{8}) \,+\, S_8(z_{1},z_{8}, \dots , z_{7},z_{6})\, .
\ee
The {\it l.h.s.} must be cyclically symmetric in $z_{i} \to z_{i-1}$. To guarantee it
we must impose the flip symmetry \eqref{flip-def} on $S_8$. When applied to the second term
on the {\it r.h.s.} of \eqref{R82S8-2} we find,
\be \label{R82S8-3}
\tilde \cR_{8}(z_{1},z_{2}, \dots ,z_{7},z_{8})\,=\,
S_8(z_{1},z_{2}, \dots , z_{7},z_{8}) \,+\, S_8(z_{8},z_{1}, \dots , z_{6},z_{7})\, ,
\ee
which automatically gives a cyclically symmetric combination, even though $S_8$ individually are
not required to have it.
We can now divide $S_8$ into two parts, so that,
\bea
S_8(x_2,x_4,x_6,x_8) &=& \frac{1}{2}\cR_{8}(z_{1},z_{2}, \dots ,z_{8}) \,+\,
T_8(x_2,x_4,x_6,x_8) \, , \label{4.5}\\
S_8(x_1,x_3,x_5,x_7) &=& \frac{1}{2}\cR_{8}(z_{1},z_{2}, \dots ,z_{8}) \,+\,
T_8(x_1,x_3,x_5,x_7) \, .\label{4.6}
\eea
$T_8$ denotes an additional contribution to $S_8$, which is not determined by the amplitude $\cR_{8}$.
To ensure that $T_8$'s indeed do not appear in \eqref{R82S8} we require that
\be
T_8(x_2,x_4,x_6,x_8) \,+\, T_8(x_1,x_3,x_5,x_7) \,=\, 0\, .
\ee
This condition is  guaranteed by the flip symmetry of $T_8$ together with the {\it anti}-symmetry under $z_{i} \to z_{i+1}$,
\be \label{T8syms}
T_8(x_1,x_3,x_5,x_7) \,=\, T_8(x^f_1,x^f_3,x^f_5,x^f_7) \,=\, - T_8(x_2,x_4,x_6,x_8)\, .
\ee
The entire $S_8$ can be constructed using the method of \cite{HK3loop} as we explain in Appendix B.
In particular, the contributions to the amplitude $\cR_{8}$ are constructed using $f^{+}$ functions and the
additional contributions -- to $T_8$ -- are constructed from $f^{-}$
functions, cf~\eqref{S8B} . In particular, $T_8^{(3)} (x_2,x_4,x_6,x_8) = b_{\sigma\tau} f^{-}_\sigma(u_1) f^{-}_\tau(u_2) $
and $T_8^{(3)} (x_1,x_3,x_5,x_7) =  - b_{\sigma\tau} f^{-}_\sigma(u_1) f^{-}_\tau(u_2) $.

We now consider the next-to-minimal case $n=10$. The first line on the {\it r.h.s.} of \eqref{eq:15-l}
gives a non-trivial sum of $S_8$ contributions. These are the contributions of $\tilde \cR_8$'s and
contributions of $T_8$'s, the latter no longer cancel each other in the sum. Novel contributions at
10-points then arise from the second line on the {\it r.h.s.} of \eqref{eq:15-l}:
\be \label{relmin}
S_{10}(x_2,x_4,x_6,x_8,x_{10}) \,-\, S_{10}(x_1,x_3,x_5,x_7,x_9) \, .
\ee
To be cyclically symmetric in $z$-variables, these functions have to be anti-symmetric under the flip symmetry
(due to the relative minus sign in \eqref{relmin}).
Together with $T_8$'s these contributions from $S_{10}$'s will give precisely the vanishing part of the 10-point function,
$V_{10}$. In a similar way there will be well-defined pieces of $S_{10}$ which do not contribute to $V_{10}$ but are instead only seen by the collinear vanishing part of the 12-point amplitude $V_{12}$. This will be explained in more detail  in section~\ref{sec:higher-points}.

We now return to the general expression \eqref{eq:15-l}.
Interestingly the formula \eqref{eq:15-l} is most simply written in terms of $x$ variables
rather than $z$'s. To see that this is non-trivial, imagine rewriting
the right-hand side back in terms of $z$ variables.  We see that
rather than having arbitrary $z$ dependence, the $z$'s only appear in
each term as pairs of nearest neighbours, i.e. if a term depends on $z_i$, then
it will necessarily depend also on either $z_{i+1}$ or
$z_{i-1}$. Writing in terms of $x$'s is a short-hand way of displaying
this dependence. Furthermore, the objects $S_m$ have properties which
are similar, but nicer than the corresponding low-point amplitudes
$\tilde \cR_m$. We will now detail the properties of $S_m$ for general $m$
before we prove that our formula correctly solves the constraints
coming from collinear limits.

The $m$-point objects $S_m$ have similar properties to $S_{8}$
discussed above. Firstly, they are conformally
invariant functions of $m$ $z$-variables or equivalently $m/2$ $x$-variables
$S_m(z_1, \dots z_m)=S_m(x_2, x_4, \dots, x_{m})$ where $x_2=(z_1,z_2)$
etc. They are also symmetric under cyclic symmetry and parity up to a minus sign in $x$-variables (but not necessarily in $z$),
\be
  S_{m}(x_2,x_4,\dots x_m)\,=\,
  S_{m}(x_4,x_6,\dots,x_m,x_2)\,=\,
  (-1)^{m/2}S_{m}(x_m,x_{m-2},\dots,x_4,x_2)\ .
\ee
Furthermore, we also require that they satisfy the additional flip
(anti)-symmetry,
\begin{align} \label{antifl-def}
  S_m(x_{i_1},x_{i_2},\dots x_{i_{m/2}})\,=\,(-1)^{m/2} \, S_m(x^f_{i_1},x^f_{i_2},\dots x^f_{i_{m/2}})\ .
\end{align}

The $S_m$'s must also vanish in the collinear limit $z_m
\rightarrow z_{m-2}$ ie $x_m \rightarrow x_{m-1}$
\begin{align}
  \lim_{x_{m} \rightarrow x_{m-1}} S_m(x_2,\dots,
  x_{m-2},x_m)&=0 &\text{(collinear limits)}\ .
\end{align}
A useful and more geometrical way of saying this
\begin{align}
  S_m(x_{i},\dots,x_{j},x_k)=0 \qquad \text{if} \qquad x_j, x_k
  \text{ become light-like separated}\ .
\end{align}

Finally $S_m$  must also vanish in the multi-collinear limits, where $(p+1)$ consecutive momenta become
collinear, $z_m,z_{m-2},\dots, z_{m-p+2} \to z_{m-p}$, or $x_m
\rightarrow x_{m-1},\
x_{m-2}\rightarrow x_{m-3},\  \dots,\  x_{m-p+2} \rightarrow
x_{m-p+1}$ for $p=2,4,\dots m-4$ ie
\begin{align}
  S_m(x_{i},x_j\dots,x_{k})=0 \qquad
  \begin{array}{l}
\text{if any set of $2,3,\dots$ or $(m/2-2)$
    consecutive points}\\ \text{become mutually light-like separated.}
\end{array}
\label{eq:12}
\end{align}
In other words
we require that the $S$-functions vanish in {\em all} allowed {\em
  multi-}collinear limits.  By ``allowed'' here we mean that we can not
insist that $S_m$ vanishes when too many points become collinear by
conformal invariance (see Appendix A). The limit when $m/2-1$ points become collinear is
conformally equivalent to points being in generic
positions and so $S_m$ can not vanish in this
limit. Similarly when all $m/2$ points become collinear.

To show that~(\ref{eq:15-l}) is indeed the $n$-point function, we must first
prove that this expression is cyclic, that it satisfies the correct
properties under collinear limits and that it is unique. That~(\ref{eq:15-l}) is cyclically symmetric in $z$-variables
comes straight from its
definition, the (anti)-flip symmetry \eqref{antifl-def} together with cyclicity of $S_{m}$ in its $x$-arguments.
In the next subsection we argue that it behaves correctly in all
collinear limits. Then we discuss the uniqueness of the structure.

\subsection{$\tilde \cR_n$ has the  correct collinear limits}

Now consider the simplest
collinear limit, $z_n\rightarrow z_{n-2}$ ie $x_{n} \rightarrow
x_{n-1}$. Using
\begin{align}
  &\lim_{x_n \rightarrow x_{n-1}} S_m({i,j \dots k})
  =S_m(i,j, \dots k) \quad {\rm for} \quad i,j,\dots k \neq
  n-1,n \quad {\rm and}\nonumber \\
  &\lim_{x_n \rightarrow x_{n-1}} \left[S_m(i,j \dots
    k,n-1)-S_m(i,j, \dots k,n)\right] =0 \ ,\label{eq:2}
\end{align}
one can see that
\begin{align}
  \lim_{x_n \rightarrow x_{n-1}} \tilde \cR_n(z_1, \dots z_n) =
  \tilde \cR_{n-2}(z_1, \dots , z_{n-2})
\end{align}
as required under collinear limits.

To prove the correct property under multi-collinear limits we need to
 work a little harder. The multi-collinear limit, where $p+1$ edges become
 collinear is defined for even $p$ as $z_n,z_{n-2},\dots,z_{n-p+2}
\rightarrow z_{n-p}$. This is the same as pairwise limits on consecutive $x$-variables,
$x_n\rightarrow x_{n-1}, \ x_{n-2}\rightarrow
x_{n-3},\  \dots, \ x_{n-p+2} \rightarrow x_{n-p+1}$ as can be easily seen from Fig.~\ref{Fig:collinear}.
More
geometrically, we can separate all the $n$ $x$-variables into two sets:
\be
\overbrace{x_{n-p}, x_{n-p+1} \leftarrow   x_{n-p+2} , \dots, x_{n-1} \leftarrow x_n, x_1}^{\cS_{p+2}}
,
\overbrace{x_2,\dots, x_{n-p-1}}^{\cS_{n-p-2}}
\ee
In this
limit all
points in the set
$\cS_{p+2} =
\{x_{n-p},x_{n-p+1} \dots, x_1\}$ are becoming mutually light-like separated
(ie collinear), whereas the points in
the set
$\cS_{n-p-2} = \{x_2 \dots, x_{n-p-1}\}$  remain unchanged. Now the
$S$'s vanish whenever $r$  consecutive points become light-like
separated for  $r=2,3,\dots {m\over2}-2$ as discussed in~(\ref{eq:12}).
Since all the points in $\cS_{p+2}$ become
light-like separated from each other, this means that $S_m$
vanishes unless all, or all but one of the points are in $\cS_{p+2}$ or
unless, all, or all but one of the points are in $\cS_{n-p-2}$, ie
\begin{align}
  S_m(x_{i_1}, \dots x_{i_r},x_{j_1},\dots x_{j_{{\bar m}-r}})\rightarrow 0  \quad &{\rm for}
  \quad r=2, \dots, {\bar m}-2  \quad \text{and where}\notag\\
  &\{i_1,\dots i_{r}\} \in
  \cS_{m-p-2}
  \quad {\rm and}\quad  \{j_1,\dots j_{{\bar m}-r}\} \in
  \cS_{p+2}\ ,
\end{align}
where we have defined $\bar m=m/2$.
 So the
sum of $S$'s appearing in $\cR$ reduces to
\begin{align}
&\sum_{{2\leq i_1 \lhd \dots\lhd i_{{\bar m}} \leq n+1}}
  S_{m}(x_{i_1},x_{i_2}, \dots ,x_{i_{{\bar m}}})(-1)^{i_1+\dots
    i_{{\bar m}}} \qquad \longrightarrow \ \nonumber \\
   &\ \sum_{2\leq i_1 \lhd \dots\lhd i_{{\bar m}-1} \leq
      n-p-1} \ \sum_{j=i_{{\bar m}-1}+2}^{n+1}
  S_{m}(x_{i_1},x_{i_2}, \dots ,x_{i_{{\bar m}-1}},x_j)(-1)^{i_1+\dots
    i_{{\bar m}-1}}(-1)^j \nonumber \\&\ +\  \sum_{n-p\leq j_1 \lhd \dots\lhd j_{{\bar m}-1} \leq
      n+1} \ \sum_{i=2}^{j_1-2}
  S_{m}(x_i,x_{j_1},x_{j_2}, \dots ,x_{j_{{\bar m}-1}})(-1)^{j_1+\dots
    j_{{\bar m}-1}}(-1)^i\ .\label{eq:10}
\end{align}
Now consider the first term of this last expression, and in particular
focus on the sum over $j$. We have that
\begin{align}
&  \sum_{j=i_{{\bar m}-1}+2}^{n+1}
  S_{m}(x_{i_1},x_{i_2}, \dots ,x_{i_{{\bar m}-1}},x_j)(-1)^j \notag \\ =&  \sum_{i_{\bar m}=i_{{\bar m}-1}+2}^{n-p-1}
  S_{m}(x_{i_1}, \dots,x_{i_{\bar m}})(-1)^{i_{\bar m}} + \sum_{j=n-p}^{n+1}
  S_{m}(x_{i_1}, \dots ,x_{i_{{\bar m}-1}},x_{j})(-1)^{j} \notag \\
=& \sum_{i_{\bar m}=i_{{\bar m}-1}+2}^{n-p-1}
S_{m}(x_{i_1}, \dots,x_{i_{\bar m}})(-1)^{i_{\bar m}} +
  S_{m}(x_{i_1}, \dots ,x_{i_{{\bar m}-1}},x_{n-p})-
  S_{m}(x_{i_1}, \dots ,x_{i_{{\bar m}-1}},x_{n+1})
\end{align}
where in the last equality we have used the fact that $x_j$ is one of
the vertices becoming collinear, and in the limit  $x_n\rightarrow x_{n-1}, \ x_{n-2}\rightarrow
x_{n-3},\  \dots, \ x_{n-p+2} \rightarrow x_{n-p+1}$, thus the
alternating sum collapses to the two boundary cases. Inserting this
back into~(\ref{eq:10}) and using cyclicity, we can include this most
succinctly by including the end-points $n-p$
and $n+1=1$ in the sum to rewrite  the first term on the
right-hand side of~(\ref{eq:10}) in the suggestive form
\begin{align}
  \sum_{1\leq i_1 \lhd \dots\lhd i_{{\bar m}} \leq n-p} S_{m}(x_{i_1},x_{i_2}, \dots ,x_{i_{{\bar m}}})(-1)^{i_1+\dots
    i_{{\bar m}}}\ .\label{eq:13}
\end{align}

So we have massaged the first term on the {\it r.h.s.} of~(\ref{eq:10}) into a nice form.
 The second term in~(\ref{eq:10}), despite its similarity to the
first term looks trickier to manipulate into something pleasant, since
instead of 1 point becoming collinear ${m/2-1}$ of the points are
becoming collinear. However
at this point we can make use of the fact (used in~\cite{OPE}) that the collinear limit we are performing is
conformally equivalent to a different multi-collinear limit, in which instead of the
points in $\cS_{p+2}$ becoming collinear and the points in
$\cS_{n-p-2}$ remaining unchanged, we have the converse: the points in
$\cS_{p+2}$ remain unchanged and the points in  $\cS_{n-p-2}$ become
collinear. With this observation we see that in this conformally
equivalent setting, only the point $x_i$ is
becoming collinear and the points $x_{j}$ remain unchanged.  We can
then perform similar manipulations to those leading to~(\ref{eq:13})
on the second term on the {\it r.h.s.} of~(\ref{eq:10}) to obtain the final result
\begin{align}
&\sum_{\substack{2\leq i_1 \lhd \\\dots\\\lhd i_{{\bar m}} \leq n+1}}
  S_{m}(x_{i_1}, \dots ,x_{i_{{\bar m}}})(-1)^{i_1+\dots
    i_{{\bar m}}}\notag \\  \rightarrow &  \ \sum_{\substack{1\leq i_1 \lhd \\\dots\\\lhd i_{{\bar m}} \leq n-p}} S_{m}(x_{i_1}, \dots ,x_{i_{{\bar m}}})(-1)^{i_1+\dots
    i_{{\bar m}}}+\sum_{\substack{n-p-1\leq j_1 \lhd \\\dots\\\lhd j_{{\bar m}} \leq n+2}} S_{m}(x_{j_1}, \dots ,x_{j_{{\bar m}}})(-1)^{j_1+\dots
    j_{{\bar m}}}\ .
\end{align}
Now this is true for any value of $m$ and since our general
formula for the amplitude~(\ref{eq:15-l}) is made from such structures
as these
we conclude that in the multi-collinear limit
\begin{align}
 \tilde  \cR_n \rightarrow \tilde \cR_{n-p} + \tilde \cR_{p+4}\ ,
\end{align}
precisely as we expect.

\subsection{Discussion of the result}

So~(\ref{eq:15-l}) gives a solution of the constraints from collinear
limits. How  unique is this solution? To examine this question,
imagine that the formula~(\ref{eq:15-l}) failed to give the correct result
for $\tilde \cR_n$ at $n$-points (but succeeded below this
point). Then consider the difference between the prediction
from~(\ref{eq:15-l}) and $\tilde \cR_n$,
$\tilde \cR_n-\tilde \cR^{(\ref{eq:15-l})}_n$. Since both obey the
same collinear limits, this is an $n$-point function which vanishes in
all allowed collinear limits. So one might expect that we can always absorb
this into the collinear vanishing object $S_n$. However this not quite
as straightforward as it first appears. In the following subsections 
we will argue, first by considering explicit special cases, and then
outlining the general case, that we can always 
absorb the collinear vanishing piece into $S_n$ and $S_{n-2}$. This
means that (\ref{eq:15-l}) would indeed give the unique result if we
allowed $S_n$ for all $n$.   


We have, however also restricted the number of collinear vanishing
functions so that in particular $S_m^{(\ell)}\,=\, 0$ for $m>4 \ell$.
So we now focus on the question of why $S_n$, and hence the
collinear vanishing part of $\tilde \cR_n$ should be restricted in
this way.
The point is simply that one can not write down a collinear vanishing, conformally invariant  $\ell$-loop
function beyond $4\ell$-points. This was argued in~\cite{HK3loop} and 
for completeness we briefly recast the argument here. It is based on examining the symbol of $S_m$. The central assumption of \cite{HK3loop}
was that the basis of the symbol (in 2d kinematics) is made out of simple cross-ratios $u_{ij}$.
These cross-ratios have a clear and simple behaviour in two collinear limits, those associated with the edges $i$ or $j$.
Specifically,
$\log u_{ij} \to 0$ when either $z_{i+1}\to z_{i-1}$, or $z_{j+1}\to z_{j-1}$. Thus, the presence of
$u_{ij}$ in the symbol of $S_m$ makes it vanish in the collinear limits associated with the edges $i$ or $j$.
To make sure that $S_m$ vanishes in all possible collinear limits, its symbol must contain $u_{ij}$'s for
all pairs of edges. At $\ell$-loops, there are $2\ell$ tensor products of $u_{ij}$'s in the symbol, and they
can connect maximally $4\ell$ different edges. This means that collinearly vanishing functions exist
only up to $m=4\ell$ points.
Thus the formula~(\ref{eq:15-l}) gives a unique uplift.

In the next section we will extend this
analysis to non-MHV amplitudes and obtain similar conclusions.  In
Section~{\bf \ref{sec:tree-level-nmhv}} we will consider the tree-level NMHV
amplitudes.  They evade our conclusions by not manifestly having the
correct collinear behaviour (and indeed they are not manifestly cyclic
either). They only have these properties after taking into account
special linear identities. We believe  this is  special to tree-level and that at
loop level the only solution is~(\ref{eq:15-l}).

\subsection{Special cases}

We first look again at the $n$-point 2-loop result of~\cite{HK2loop}.
At 2-loops, inserting the 8-point result for $S_8^{(2)}$
\begin{align}
  S^{(2)}_{8}(z_1,\dots z_8)=-{1\over 4}\log\left(u_{17;53}\right)\log\left(u_{31;75}\right)\log\left(u_{28;64}\right)\log\left(u_{42;86}\right)
\end{align}
into~(\ref{eq:15-l})
\begin{align}\label{eq:15-2}
  \tilde \cR^{(2)}_{n}(z_{1},z_{2}, \dots ,z_{n}) &= \sum_{1\leq
    i_1\lhd i_2\lhd i_3\lhd i_4\leq n} S^{(2)}_{8}(x_{i_1},x_{i_2},
  x_{i_3},x_{i_4})(-1)^{i_1+\dots
    i_{4}}\ ,
\end{align}
and rewriting in terms of the basis $u_{ij}$s correctly reproduces the
form of the two-loop result in~(\ref{Rn2-result}).

Next, at 3 loops
the formula~(\ref{eq:15-l}) for any number of points contains
essentially only three independent terms. It reduces to
\begin{align}\label{eq:15}
  \tilde \cR^{(3)}_{n}(z_{1},z_{2}, \dots ,z_{n}) &= \sum_{1\leq
    i_1\lhd i_2\lhd i_3\lhd i_4\leq n} S^{(3)}_{8}(x_{i_1},x_{i_2},
  x_{i_3},x_{i_4})(-1)^{i_1+\dots
    i_{4}}\ \nonumber\\
 &+ \sum_{{1\leq i_1 \lhd \dots\lhd i_5 \leq n}}
  S^{(3)}_{10}(x_{i_1},x_{i_2}, \dots ,x_{i_{5}})(-1)^{i_1+\dots
    i_{5}}
    \nonumber\\
  &+ \sum_{{1\leq i_1 \lhd \dots\lhd i_6 \leq n}}
  S^{(3)}_{12}(x_{i_1},x_{i_2}, \dots ,x_{i_{6}})(-1)^{i_1+\dots
    i_{6}}\ ,
\end{align}
where the multi-collinearly-vanishing function $S_{m}$ are
constructable with methods of \cite{HK3loop} as we will now
demonstrate. We will show
that the general formula~(\ref{eq:15-l}) correctly reproduces the
10-point result of~(\ref{eq:16}) and gives the entire
collinear vanishing term $V_{10}$.

\subsubsection{$S_{10}$ contribution to $V_{10}$}

We first consider the $S_{10}$ collinear vanishing contribution to
$\cR_{10}$. The building blocks are collinear vanishing functions ($f_1,f_2,f_3$)
of even and odd cross-ratios, derived in~\cite{HK3loop} and listed
in~\eqref{eq:B-15}. Since $f_1$ and $f_2$  give 5 independent functions via cyclic permutations of their
arguments, whereas $f_3$ is already cyclically symmetric, giving only
1 
independent function, we have in total 11
functions.

Let us now rewrite these functions in a
basis which diagonalises the action of the cyclic group\footnote{Here
  we mean the cyclic group which acts separately on the even and odd
  variables, so in this case it is $C_5$},
\begin{align}
  f_1^{(k)}(z_1,z_3,z_5,z_7,z_9)&:=\sum_{j=1}^5 f_1(u_{2j},u_{2j+2},u_{2j+4})e^{2 \pi i
    jk/5}\qquad k=0\dots 4\nonumber\\ 
  f_2^{(k)}(z_1,z_3,z_5,z_7,z_9)&:=\sum_{j=1}^5 f_2(u_{2j},u_{2j+2},u_{2j+4})e^{2 \pi i
    jk/5}\qquad k=0\dots 4\nonumber\\ 
  f_3^{(0)}(z_1,z_3,z_5,z_7,z_9)&:=f_3(u_2,u_4,u_6,u_8,u_{10})
\end{align}
These new functions lie in irreducible representations of
the cyclic group, in fact they are eigenstates of the cyclic group,
\begin{align}
  f_a^{(k)}(z_3,z_5,z_7,z_9,z_1)=e^{2\pi i
    k/5}f_a^{(k)}(z_1,z_3,z_5,z_7,z_9)\ .
\end{align}
 We
also have that under parity $f^{(k)}\rightarrow f^{(5-k)}$.

Then by construction both $V_{10}$ appearing in~\eqref{eq:14} and
$S_{10}$ appearing in~(\ref{eq:15-l})
are $C_5$ and parity invariant combinations of these functions. To
obtain cyclic ($C_5$)
invariant combinations, a function carrying cyclic representation $k$
must multiply a  function carrying cyclic representation $-k$.

Let us first construct $V_{10}$. It is given by a linear combination
of 
the 12 collinear vanishing contributions to the remainder function
listed in~(\ref{eq:BlastB}). These are now written as
\begin{align}
  &f_a^{(k)}(z_\text{odd})f_b^{(-k)}(z_\text{even})+e^{-2\pi i
    k/5}f_b^{(-k)}(z_\text{odd})f_a^{(k)}(z_\text{even}) \quad + \quad
  a\leftrightarrow b \ ,\label{eq:21}\\
  &f_3^{(0)}(z_\text{odd})f_a^{(0)}(z_\text{even})+
  f_a^{(0)}(z_\text{odd})f_3^{(0)}(z_\text{even})\qquad
  \qquad a=1,2,3\ ,\label{eq:19a}
\end{align}
where $z_\text{odd}:=z_1,z_3,z_5,z_7,z_9$ and
$z_\text{even}:=z_2,z_4,z_6,z_8,z_{10}$. In the first equation we have
$a,b=1,2$ and $k=0,1,2$, thus it gives 9 independent functions, in the
second equation $a=1,2,3$ giving 3 more. Clearly these 12 functions
are simple recombinations of the 12 functions
in~(\ref{eq:BlastB}). This is what we have for $V_{10}$.

Let us compare this with the construction of $S_{10}$. These are
constructed from the same building block functions, with an additional
constraint that $S_{10}$ must be antisymmetric under flip symmetry. They are given as
\begin{align}
S_{10}(x_2,x_4,x_6,x_8,x_{10}) \ni &f_a^{(k)}(z_\text{odd})f_b^{(-k)}(z_\text{even})-f_b^{(-k)}(z_\text{odd})f_a^{(k)}(z_\text{even})\nonumber\\
  &+f_b^{(k)}(z_\text{odd})f_a^{(-k)}(z_\text{even})-f_a^{(-k)}(z_\text{odd})f_b^{(k)}(z_\text{even})\ .
\end{align}
Non vanishing contributions arise from  $k=1,2$ and $a,b=1,2$, so we
have 6 contributions in total.
Note in particular that the invariant representation $k=0$ drops out
here. Now, the contribution from $S_{10}$'s to $\cR_{10}$, dictated by
the $S$-formula~(\ref{eq:15-l}), 
\begin{align}
 \tilde \cR_{10} \ni  S_{10}(x_2,x_4,x_6,x_8,x_{10})-S_{10}(x_1,x_3,x_5,x_7,x_{9})
\end{align}
is
\begin{align}
  &(1-e^{2\pi
    ik/5})\Big(f_a^{(k)}(z_\text{odd})f_b^{(-k)}(z_\text{even})+e^{-2\pi i
    k/5}f_b^{(-k)}(z_\text{odd})f_a^{(k)}(z_\text{even}) \quad + \quad
  a\leftrightarrow b \Big) \label{eq:20}
\end{align}
We can now see that the contribution of $S_{10}$'s to the 10-point
amplitude~(\ref{eq:21}) gives a clearly identifiable subset of the
most general collinearly vanishing contribution $V_{10}$
in~(\ref{eq:19a}). They are the same functions, simply multiplied by a
constant factor $(1-e^{2\pi
    ik/5})$ which plays no role,  except in the case $k=0$ where it
  vanishes. 

Thus we see clearly that the contribution of $S_{10}$  yields the entire collinear
vanishing part of $\cR_{10}$ {\em{except}} the pieces constructed from
the cyclically 
invariant functions $f_a^{(0)}$. We will now see how these missing
building blocks are correctly filled in by contributions from  $S_8$
or more precisely $T_8$.

\subsubsection{$S_8$ contribution to $V_{10}$}
\label{sec:s_8--contribution}

Now consider the contribution of $S_8$ to $\cR_{10}$. Following
Section~{\bf \ref{sec:general-formula-n}} we split $S_8$ into $\cR_8$
and $T_8$ parts~\eqref{4.5}-\eqref{4.6}. The
role of $\cR_8$ is completely clear. It is the
8-point amplitude and furthermore it contributes to the collinearly
non-vanishing part of all higher point amplitudes. The first
contribution of $T_8$ however arises only at 10-points where it
contributes to the collinearly vanishing part of the answer.
Here we wish to trace through the $T_8$  contribution to $V_{10}$.

From~\eqref{S8B}
We have
\bea 
T_8 (x_2,x_4,x_6,x_8) \,&=&\, \sum_{\sigma,\tau}\,
\, b_{\sigma\tau} f^{-}_\sigma(u_1) f^{-}_\tau(u_2)
\eea
where $b_{\sigma\tau}=b_{\tau\sigma}$ and the functions $f^-_\sigma$,
$\sigma=1,2,3$,   are
listed in~(\ref{eq:4cB2}). These functions $f^-_{\sigma}$ are all
weight 3. It turns out that contributions of $T_8$ of the form  (weight 2) $\times$ (weight 4)  
vanish at all points. We will discuss this point further at the end of this subsection.     In terms of the $z$-variables these
functions satisfy the following property (cf~\eqref{eq:9pm}),
\begin{align}
  f^{-}_\sigma(z_3,z_5,z_7,z_9)=-f^{-}_\sigma(z_1,z_3,z_5,z_7)\ ,
\end{align}
 ie they are invariant with an alternating sign under cyclic
 symmetry.

 Inserting $T_8$  into the S-formula
 \begin{align}
   \tilde \cR_{10} &\ \ni\   \sum_{1\leq
    i_1\lhd  i_2\lhd  i_3\lhd  i_4\leq 10} S_{8}(x_{i_1},x_{i_2},
  x_{i_3},x_{i_4})(-1)^{i_1+i_2 +i_3 +   i_{4}}\nonumber \\
  &\ \ni\   \sum_{1\leq
    i_1\lhd  i_2\lhd  i_3\lhd  i_4\leq 10} T_{8}(x_{i_1},x_{i_2},
  x_{i_3},x_{i_4})(-1)^{i_1+i_2 +i_3 +   i_{4}}
 \end{align}
and performing the sum we have
\begin{align}\label{eq:23}
\tilde \cR_{10} \ \ni \ \sum_{\sigma,\tau} b_{\sigma,\tau} F_\sigma(z_1,z_3,z_5,z_7,z_9) F_\tau(z_2,z_4,z_6,z_8,z_{10})
\end{align}
where 
\begin{align}
  &F_\sigma(z_1,z_3,z_5,z_7,z_9)\label{eq:22}\\
  &=
f^-_\sigma(z_1,z_3,z_5,z_7)-f^-_\sigma(z_1,z_3,z_5,z_9)+f^-_\sigma(z_1,z_3,z_7,z_9)-f^-_\sigma(z_1,z_5,z_7,z_9)+f^-_\sigma(z_3,z_5,z_7,z_9)\nonumber\\&=
f^-_\sigma(z_1,z_3,z_5,z_7)+f^-_\sigma(z_9,z_1,z_3,z_5)+f^-_\sigma(z_7,z_9,z_1,z_3)+f^-_\sigma(z_5,z_7,z_9,z_1)+f^-_\sigma(z_3,z_5,z_7,z_9)\nonumber
\end{align}

Note that the functions $F_\sigma$, although constructed from four-point building blocks,  are in fact cyclically
invariant 5-point
functions. Furthermore, inspection of the {\em{r.h.s.}}
of~(\ref{eq:22}) shows that  they also vanish in collinear limits.
Thus we see that the {\em{r.h.s}} of~(\ref{eq:23}) corresponds
precisely to $k=0$ contributions to $V_{10}$ in
(\ref{eq:21}),(\ref{eq:19a}). We have six contributions 
\begin{align}
  F_1F_1,\ F_1F_2+F_2F_1,\ F_1F_3+F_3F_1,\ F_2F_2,\ F_2F_3+F_3F_2,\
  F_3F_3
\end{align}
and these are the six previously missing contributions in $V_{10}$,
not accounted by $S_{10}$, of the previous subsection.

One obvious question is what happens if we use (weight 2) $\times$
(weight 4) functions, $f^-$, to construct $T_8$. According to the above
discussion this should produce  a (weight 2) $\times$ (weight 4)
collinear vanishing contribution to $S_{10}$ which we know is not
present in $V_{10}$ giving an apparent contradiction. 
In reality it is easy to see that all such contributions vanish. There
is a unique weight 2 function $f^-(u)=\Li_2(u)-\Li_2(1-u)$ and when we
plug it into~(\ref{eq:22}) we see that  the corresponding function $F$
vanishes. When written in
terms of the symbol this identity is manifest; in terms of the
polylogarithms this becomes the equation
\begin{align}
  \Li_2(u_1)+\Li_2(u_3)+\Li_2(u_5)+\Li_2(u_7)+\Li_2(u_9) -\ 
  (u_i\leftrightarrow 1-u_i)\  =\  \text{constant}
\end{align}
which, when writing in terms of $u_1, u_5$ via the relation~\eqref{n10const}
\begin{align}
  u_3=1-u_1 u_5 \qquad u_7={1-u_5 \over 1-u_1 u_5} \qquad u_9={1-u_1 \over 1-u_1 u_5}
\end{align}
is equivalent to the famous non-trivial five-term identity for the
dilogarithm, first 
discovered by Spence in 1809. Thus as mentioned earlier no (weight 2)
$\times$ (weight 4) contributions survive in $V_{10}$ while weight
3  functions have already been accounted above.

We also note that the contributions involving weight 2 functions $f^-$ 
also disappear from $V_n$ at all higher $n$.

To summarise we have demonstrated that $T_8$ and $S_{10}$ together
generate all possible collinear vanishing 10-point
functions. And this confirms that the $S$-formula does not miss anything.

\subsection{Higher points}
\label{sec:higher-points}
This general pattern continues in a similar way to higher points. We
construct $S_{2m}$'s from the product of collinear vanishing building block functions of
even and odd $z$'s. We choose a basis of these which diagonalise the
cyclic group, and call them $f_{a}^{(k)}$ where $k$ is the
representation of the cyclic group $C_m$ and $a$ labels the inequivalent functions. Then the $S$-formula gives
the contribution
\begin{align}
  S_{2m}\ =\ &\sum_{a,b}\alpha_{ab;k} \,\Big(
  f_{a}^{(k)}(z_\text{odd})f_{b}^{(-k)}(z_\text{even})+(-1)^m
  f_b^{(-k)}(z_\text{odd})f_a^{(k)}(z_\text{even})\Big) \ + \ 
  \text{parity} 
\end{align}
where $k=0,1, \dots m$, giving the contribution to (the collinear vanishing part of) $\cR_{2m}$
of
\begin{align}
  &\sum_{a,b}\alpha_{ab;k} \, (1+(-1)^me^{2\pi i k/m}) \Big(f_{a}^{(k)}(z_\text{odd})f_{b}^{(-k)}(z_\text{even})+e^{-2\pi i
    k/m}f_b^{(-k)}(z_\text{odd})f_a^{(k)}(z_\text{even})\Big) \nonumber\\
  &\hspace{5cm} + \ 
  \text{parity}\ .
\end{align}
This yields all possible collinear vanishing $m$-point amplitudes,
simply multiplied by an irrelevant overall factor $(1+(-1)^me^{2\pi i k/m})$,
except those built from $k=0$ ($m$ odd) or $k=m/2$ ($m$ even) where
the factor vanishes. In
other words the $S_{2m}$ contribution to $\cR_{2m}$ misses out the cyclically invariant symmetric
building block functions if $m$ is odd,  or in the $m$ even case it
misses out the functions cyclically invariant up to a sign.

However, just as in Section~{\bf \ref{sec:s_8--contribution}}, these missing
contributions at $2m$-points
will arise from $2m-2$ points
and together should fill the full space of collinearly vanishing
functions $V_{2m}$.

Thus the $S$-formula~(\ref{eq:15-l}) gives an explicit formula for the
$n$-point amplitude, in terms of collinearly vanishing objects $S_m$
once they have been constructed, and in this paper we have shown how
to do that using the method of~\cite{HK3loop}.

\section{Collinear uplift of $n$-point $N^k$MHV amplitudes }

\label{sec:collinear-uplift-n}

The general formula for lifting MHV amplitudes to higher points looks very general and immediately suggests generalisation to $N^k$MHV superamplitudes. To do so we will need to examine odd superspace variables in 2d and the form of the collinear limit.

As discussed in Section~{\bf \ref{sec:two}},  superamplitudes can be written in chiral superspace depending on superspace coordinates $X_i=(x_i,\theta^{A}_i)$ where the bosonic components $x_i$
are given in terms of 2d lightcone coordinates by Eq.~\eqref{eq:6}. Examining the implications of the light-like condition for the $\theta$'s~(\ref{thetadef}) in 2d kinematics, we find that
the condition can be solved in an
analogous manner to the way we write $x$'s in terms of $z$'s, namely for the Grassmann coordinates,
$\theta$'s and $\chi$'s
\begin{align}\label{eq:theta2d}
  \theta_i^{\alpha A}=\left\{
  \begin{array}{ll}
(\chi_{i-1}^A, \chi_{i}^A)\ , \qquad &i\ \rm{even}\\
       (\chi_i^A, \chi_{i-1}^A)\ , \qquad &i\ \rm{odd}
     \end{array}\right. \ .
\end{align}
Indeed, comparing with the supertwistor in the form of Eq.~\eqref{2dstwist} we find that the $\chi$'s are precisely the odd supertwistor variables just as the $z$ were the bosonic twistors.

The general formula~(\ref{eq:15-l}) giving all $n$-point MHV amplitudes
in terms of a finite number of collinear vanishing functions
generalises immediately now to the non-MHV case. Indeed the collinear limits
$z_n \rightarrow z_{n-2}$ must be accompanied by identical limits
for the Grassmann coordinates $\chi_n \rightarrow \chi_{n-2}$. Indeed,
one has to be careful about the relevant speed at which we take
the limit. We here take the collinear limit in a supersymmetric way. More
precisely the collinear limit can be taken as a particular
superconformal transformation on the relevant vertices. The details
of this limit are given in Appendix A.

So precisely as for the MHV case  we have collinear vanishing
functions, this time  of the super-co-ordinates $S_m(X_2, X_4, \dots, X_m)$
which satisfy cyclicity and (anti-)parity in their $(X)$ arguments, flip (anti-)symmetry and collinear vanishing
property in all (allowed) collinear limits, so
\begin{align}
    S_m(X_2, X_4, \dots X_{m})=&  S_m(X_4, X_6, \dots X_{m},X_2) \nonumber\\
  =&  (-1)^{m/2} S_m(X_m, X_{m-2}, \dots X_{2}) \nonumber\\
  =&  (-1)^{m/2} S_m(X^f_2, X^f_4, \dots X^f_{m}) \nonumber\\
   \lim_{X_m \rightarrow X_{m-1}}  S_m(X_2, X_4, \dots X_m) =& 0\ .
\end{align}
Or more generally $S_m$ vanishes whenever any (allowed) number of
consecutive $X$'s become light-like separated (in the supersymmetric
sense:  $X_1=(x_1,\theta_1)$ and $X_2=(x_2,\theta_2)$ are
light-like separated if $x_{12}^2=0$ and $\theta_{12\alpha}x_{12}^{\alpha
  \dot \alpha}=0$)  i.e.
\begin{align}
  S_m(X_{i},X_j\dots,X_{k})=0 \qquad
  \begin{array}{l}
\text{if any set of $2,3,\dots$ or $m/2-2$
    consecutive points}\\ \text{become mutually light-like separated.}
\end{array}
\label{eq:12-s}
\end{align}

We note that $S_m(X_1, X_3, \dots X_{m-1})$ is a function of superspace variables,
and is not the same object as the  MHV function $S_m(x_1, x_3, \dots x_{m-1})$ with purely bosonic-variables
from the previous section. The latter however is given by the zero-th order in $\theta$ expansion
of the former. As we are talking here about superamplitudes, the $N^k$MHV label $k$ does not appear
in $\tilde \cR_{n}$ and $S_n$ in formulae below, but the $N^k$MHV
 amplitudes, $\tilde \cR_{n,k}$, will  arise as $\theta^{4k}$ components
of $\tilde \cR_{n}(X)$.

The general formula for the $n$-point amplitude is given, in exact
analogy with the MHV case, by
\begin{align}\label{eq:15-k}
  \tilde \cR^{(\ell)}_{n}(\cZ_{1},\cZ_{2}, \dots ,\cZ_{n}) =&  \sum_{{1\leq i_1\lhd \dots\lhd i_4\leq n}} S^{(\ell)}_{8}(X_{i_1},X_{i_2},
  \dots
  ,X_{i_{4}})(-1)^{i_1+\dots i_{4}}\ \\
&+\sum_{{1\leq i_1\lhd \dots\lhd i_5\leq n}} S^{(\ell)}_{10}(X_{i_1},X_{i_2},
  \dots
  ,X_{i_{5}})(-1)^{i_1+\dots i_{5}}\ \nonumber\\ 
  &+\sum_{{1\leq i_1\lhd \dots\lhd i_6\leq n}} S^{(\ell)}_{12}(X_{i_1},X_{i_2},
  \dots
  ,X_{i_{6}})(-1)^{i_1+\dots i_{6}}\ \nonumber\\  &+ \dots \nonumber\\
  &+ \sum_{{1\leq i_1 \lhd \dots\lhd i_{m_{\text{max}}/2} \leq n}}
  S^{(\ell)}_{m_{\text{max}}}(X_{i_1},X_{i_2}, \dots ,X_{i_{m_{\text{max}}/2}})(-1)^{i_1+\dots
    i_{m_{\text{max}}/2}}\ .\notag
\end{align}
One can easily verify that~(\ref{eq:15-k}) gives a formula with
the correct properties under collinear limits. Indeed for the multi-collinear
limit in which superspace points in the set $\cS_{p+2} =
\{X_{n-p},X_{n-p+1} \dots, X_1\}$ become light-like separated (in the
supersymmetric sense)
from all other points in $\cS_{p+2}$ (i.e. collinear) whereas the points in
the set
$\cS_{n-p-2} = \{x_2 \dots, x_{n-p-1}\}$  remain
unchanged. Importantly this limit can be described by performing a
conformal transformation on the points in $\cS_{p+2}$ (see Appendix A). In this limit
one can see that
\begin{align}
\tilde  \cR_n \rightarrow \tilde \cR_{n-p} + \tilde \cR_{p+4}\ ,
\end{align}
exactly as required~\eqref{superRcolllog}.
The proof follows by direct analogy to the arguments in the MHV case around~(\ref{eq:2}).

Thus the only
question is how many $S$'s are there, i.e. what is $m_{\text max}$. This
will depend on the loop level $\ell$ and the order in $\chi$-expansion, i.e. the value of $k$.
Based on the MHV bound, $m_{\text MHV}\le 4\ell$ and the $\overline{Q}$-equation
of Ref.~\cite{ch-he,Bullimore:2011kg} which related N$^k$MHV amplitudes at $\ell$-loops to
$\overline{Q}\cdot$N$^{k-1}$MHV amplitudes at $(\ell+1)$-loops, one could expect that
$m_{\text max}=4(\ell+k)$.

\section{Tree-level NMHV amplitude}
\label{sec:tree-level-nmhv}
In this section we  reduce the known $n$-point tree-level NMHV superamplitudes down to 2d kinematics. This is a non-trivial procedure,
since each term diverges in 2d kinematics and only certain combinations are finite.

In full 4d kinematics, the tree-level NMHV amplitude is~\cite{Drummond:2008bq,Drummond:2008vq}
\begin{align}\label{eq:7}
  \cR_{n;1}^{\text tree}=\frac{1}{2} \sum_{i,j} \left[ 1,i-1,i,j-1,j \right]\ .
\end{align}
where the 5-brackets (which are totally anti-symmetric in their
arguments) can be written in momentum supertwistors~(\ref{stwistordef}) as~\cite{Mason:2009qx}
\begin{equation}
\left[ i,j,k,l,m \right]= \frac{\delta^{0|4} \left( \chi^{i} \vev{jklm} + \mathrm{cyclic} \right)}{\vev{ijkl}\vev{jklm}\vev{klmi}\vev{lmij}\vev{mijk}}\ . \label{MomTwiR}
\end{equation}
The $4$-brackets $\vev{ijkl}$ are defined in~(\ref{eq:3}) and as
mentioned there, in 2d kinematics these vanish unless there are
precisely 2 even particles and 2 odd particles. This clearly can not
be the case for all five 4-brackets in the denominator, and so we
conclude that the 5-bracket inevitably diverges in 2d. None-the-less
it must be that the 2d kinematics should lead to sensible amplitudes,
so~(\ref{MomTwiR}) should be rewritable in terms of some finite
combinations of 5-brackets. At this point we notice that in fact the
poles which diverge in 2d are in fact spurious poles which can not be
present in the amplitude itself. Guided by this insight and the
analysis of spurious poles given in~\cite{Korchemsky:2009hm} we are
able to find simple combinations of 5-brackets which are finite in
4d. Furthermore, these combinations of 5-brackets factorise into two
3-brackets in  2d kinematics:
\begin{align}
   \tilde R(i,j,k):=
  &  \vevv{i,j-1,j,k-1,k}  + \vevv{j,k-1,k,i-1,i}+\vevv{k,i-1,i,j-1,j}\notag\\&=\vevv{i,j,k}\vevv{i-1,j-1,k-1} \notag \qquad
  \text{ ($i,j,k$ all even/odd)}\notag\\[10pt]
  \tilde R(i,j,k):=
  &-\vevv{i,j-1,j,k-1,k}  -
  \vevv{j,k-1,k,i-1,i}+\vevv{k-1,i-1,i,j-1,j}\notag\\
  &=\vevv{i,j,k-1}\vevv{i-1,j-1,k}\notag \qquad  \text{($i,j$ even/odd
    ; $k$ odd/even)}\\[10pt]
   \tilde R(i,j,k):=
  & \vevv{i,j-1,j,k-1,k} - \vevv{j-1,k-1,k,i-1,i}-\vevv{k-1,i-1,i,j-1,j}\notag \\& =\vevv{i,j-1,k-1}\vevv{i-1,j,k} \qquad \text{(  $i$ even/odd ; $j,k$ odd/even)}\ .\label{eq:4}
\end{align}
Here on the right-hand side we have used 3-brackets, the natural
analogue of the 5-bracket in 2d kinematics, an invariant of
$SL(2|2)$,
defined as
\begin{align}
  \vevv{ijk} = \frac{\delta^{0|2} \left( \chi^{i} \vev{jk} + \chi^{j} \vev{ki}+\chi^{k} \vev{ij}\right)}{\vev{ij}\vev{jk}\vev{ki}}\ .
\end{align}

In actual fact, as mentioned earlier, at least at NMHV level, we do not need to reduce the internal $SU(4)$ group, but can perfectly well keep the full $\chi$ structure. Ie we do not need to reduce the 4-component $\chi$'s to 2-component $\chi$'s as in~(\ref{2dstwist}). In this case the above 3-brackets would simply have two antisymmetric $SU(4)$ indices $\vev{ijk}^{AB}$ which would be contracted with an $\epsilon_{ABCD}$ in~(\ref{eq:4}). Nevertheless for simplicity we stick to the reduced version.

Notice that just as the 5-brackets satisfy a 6-term identity
\begin{equation}
\left[ijklm \right] + \left[jklmn \right] + \left[klmni \right] + \left[lmnij \right] + \left[mnijk \right] + \left[nijkl \right] = 0 \label{6termID}\ ,
\end{equation}
(and indeed one can use this identity to rewrite~(\ref{eq:4}) in an alternative way) so the 3-brackets satisfy a simple 4-term identity:
\begin{align}\label{eq:1}
  \vevv{ijk}- \vevv{jkl}+ \vevv{kli}- \vevv{lij}=0\ .
\end{align}
This can be quickly checked by considering $\chi$-components and
using Schouten identities.

Now let us reduce the NMHV tree-level amplitudes to 2d
kinematics. Consider first of all the first non-trivial case, the
6-point amplitude. This is
\begin{align}
  \cR_{6;1}^{\text tree}=\frac{1}{2} \Big(
  \vevv{13456}+\vevv{12356}+\vevv{12345}\Big) = \frac12
  \,\tilde R(1,3,5) = \frac12 \,\vevv{135}\vevv{246}\ .\label{eq:8}
\end{align}
Here the first equality comes directly from the general
formula~(\ref{eq:7}), and the second and third from~(\ref{eq:4}).

By considering higher points, in particular we looked at 8- and
10-points in great detail. Due to the large number of identities, it
is not clear which is the best way of representing any amplitude at
low points.  However gradually a general picture begins to emerge and
we obtain a simple formula for the $n$-point NMHV tree-level
amplitude in 2d kinematics in terms of 3-brackets. The result can be
written
\begin{align}
  \cR_{n;1}^{\text tree}=\sum_{4\leq j\lhd k\leq n}  {1\over 2}\tilde{R}\left( 2,j,k \right) \left( -1
\right)^{j+k}\ .\label{eq:17}
\end{align}
which at 6-points correctly reproduces~(\ref{eq:8}).

Let us then compare this NMHV  tree-level result with our general
result for loop level superamplitudes, given in
formula~(\ref{eq:15-k}). First of all we see that the formulae are
strikingly similar with the same type of alternating sum. The
tree-level formula starts at 6 points however whereas~(\ref{eq:15-k})
starts at 8-points.
Looking
closer we see that the main
difference is that in the tree-level formula~(\ref{eq:17}) only two  out of the
three indices are summed over, the first index remaining fixed. This
does not look cyclically invariant and indeed verification of
cyclic invariance requires the implementation of non-trivial linear
identities between the six-point $\tilde{R}\left( i,j,k \right)$ at
different points.
We can of course make it manifestly cyclically symmetric by adding
together cyclic terms to give
\begin{align}
  \cR_{n;1}^{\text tree}=\sum_{i\lhd j\lhd k \lhd i}  {1\over 2n}\tilde{R}\left( i,j,k \right) \left( -1
\right)^{j+k}\ .\label{eq:17b}
\end{align}
This has a form very similar to the general
$S$-formula~(\ref{eq:15-k}), the difference is the
appearance of the
rather asymmetric looking
$(-1)^{j+k}$ instead of the more symmetric  $(-1)^{i+j+k}$ one would
expect. Indeed, imagine extending the $S$-formula to $m=6$ to give
\begin{align}\label{eq:15-k-6}
    \sum_{{1\leq i\lhd j\lhd k\leq n}} S_{6}(i,j,k)(-1)^{i+j+k}
\end{align}
with a $(-1)^{i+j+k}$ factor.

So the question remains, how does the tree-level NMHV formula get
round this obstacle? The answer is that the $S$-formula is derived to
obey {\em manifest} cyclicity and {\em manifest} collinear limits. The NMHV
tree-level formula does not satisfy these requirements, but instead
only satisfies cyclicity {\em after taking into account non-trivial linear identities}.

For example, first  consider taking the triple/soft
collinear limit $\cZ_n \rightarrow \cZ_{n-2}$ (ie $X_n \rightarrow X_{n-1}$) on the tree-level NMHV
expression~(\ref{eq:17}). This gives
\begin{align}
\sum_{4\leq j\lhd k\leq n}  {1\over 2}\tilde{R}\left( 2,j,k \right) \left( -1
\right)^{j+k} \ \underset{X_n \rightarrow X_{n-1}}{\longrightarrow}\
&\sum_{4\leq j\lhd k\leq n-2}  {1\over 2}\tilde{R}\left( 2,j,k \right) \left( -1
\right)^{j+k} + {1\over 2}\tilde{R}\left( 2,n-2,n\right)\notag
\end{align}
correctly reproducing the collinear limit $\cR_{n;1}\rightarrow
\cR_{n-2;1} +\cR_{6;1}$ manifestly. On the other hand if we instead
perform the limit $\cZ_{n-1} \rightarrow \cZ_{n-3}$ (ie $X_{n-1} \rightarrow
  X_{n-2}$) on the tree-level NMHV expression~(\ref{eq:17}) we get
  \begin{align}
&\sum_{4\leq j\lhd k\leq n}  {1\over 2}\tilde{R}\left( 2,j,k \right) \left( -1
\right)^{j+k} \\\notag\ \underset{X_{n-1} \rightarrow X_{n-2}}{\longrightarrow}\
&\sum_{\substack{4\leq j\lhd k\leq n\\j,k \neq n-1,n-2}}  {1\over 2}\tilde{R}\left( 2,j,k \right) \left( -1
\right)^{j+k} \\&+ {1\over 2}\left(\tilde{R}\left( 2,n-3,n-1\right)-\tilde{R}\left( 2,n-3,n\right)+\tilde{R}\left( 2,n-2,n\right)\right)\notag
\end{align}
This also correctly reproduces the collinear limit $\cR_{n;1}\rightarrow
\cR_{n-2;1} +\cR_{6;1}$ but only after taking into account the linear
identity (coming from~(\ref{eq:1}))
\begin{align}
  \tilde{R}\left( 2,n-3,n-1\right)-\tilde{R}\left(
    2,n-3,n\right)+\tilde{R}\left( 2,n-2,n\right) = \tilde
  R(1,n-3,n-1)\ .
\end{align}

\bigskip

\section*{Acknowledgements}

We would like to thank Matthew Bullimore for interesting discussions.
VVK gratefully acknowledges the support of the Wolfson Foundation and
 the Royal Society. PH and VVK acknowledge support from
STFC through the Consolidated Grant number ST/J000426/1 and the IPPP grant. TG acknowledges support
from an EPSRC studentship.



\appendix


\section{Collinear limits and (super)conformal transformations}

The reason for the very simple form of the collinear
factorisation of reduced amplitudes under the $m+1$ collinear limit
comes from universal collinear factorisation of superamplitudes,
combined with (dual) superconformal symmetry. Applying the $m+1$
collinear limit on a $(m+4)$-point reduced amplitude gives the 4-point
superamplitude (which is simply 1 for the reduced superamplitude)
multiplied by the splitting superamplitude. On the other hand as we
shall show now,
performing the $m+1$ collinear limit on the $m+4$ point superamplitude
can be achieved via a superconformal transformation. Indeed this
superconformal transformation will become the definition of the
collinear limit, defining precisely the relative speed with which the
fermionic coordinates approach collinearity compared to the bosonic
variables. We will give collinear limits in terms of superconformal transformations  for the case of interest in this paper
only, namely in 2d kinematics, since the discussion is particularly
simple here: we discuss the superconformal group $SL(2|2)$ acting on
unconstrained variables $(z,\chi)$. The bosonic case is simply
the well-known M\"obius transformation.
The general 4d bosonic case was
discussed in~\cite{OPE} where it was related to the family of
conformal transformations preserving  a light-like square and the
generalisation of this to the superspace case should follow.

So we start with an $(m+4)$-point reduced superamplitude
$\cR_{m+4}(\cZ_1,\dots \cZ_{m+4})$ and wish to perform the $m+1$ collinear
limit on this. To this effect we want to send $z_{m+4}, z_{m+2}, \dots
z_{6} \rightarrow z_{4}$ and similarly $\chi_{m+4}, \chi_{m+2}, \dots
\chi_{6} \rightarrow \chi_{4}$. In particular all odd-point variables are
unchanged and we do not act on them (in 2d kinematics they are acted
on via a separate $SL(2|2)_+$ which we can choose to be the identity) but more importantly $z_2$ and
$\chi_2$ are also unchanged. In
other words we wish to find an $SL(2|2)_-$ transformation (or more
 precisely  family of transformations) which keeps $z_2,\chi_2$ fixed
 whilst all other $z \rightarrow z_4$ and all other $\chi \rightarrow
 \chi_4$.

 We can find precisely such a transformation. We use
 standard coset techniques to implement the $SL(2|2)$
 transformations. For example, the conformal part of $SL(2|2)$ acts as
 follows
 \begin{align}
   z\rightarrow {az+b \over c z+d} \, ,\qquad    \chi \rightarrow
   {\chi \over c z+d} \ .
 \end{align}
  We first use this to send $z_2 \rightarrow 0, \ z_4 \rightarrow \infty$
 and $\chi_2, \chi_4 \rightarrow 0$. At this point there is a simple
 family of transformations keeping these points fixed ($b=c=0\,,\
 d=1/a$), so that $z\rightarrow a^2
 z$, $\chi \rightarrow a \chi$ with $a$ parametrising a family of
 conformal transformations, and $a\rightarrow 0$ corresponding to the
 collinear limit. Finally, transforming back to the original coordinates we
 thus construct the explicit conformal transformation implementing our
 collinear limit as
 \begin{align}
   z\rightarrow {z_2\,a^2(z-z_4)-z_4(z-z_2) \over
     a^2(z-z_4)-(z-z_2)}\qquad \chi \rightarrow {a \,\chi \, (z_{4}-z_2)
     +(1-a)\big[\,a\,\chi_2\,(z-z_4)+\chi_4\,(z-z_2)\big] \over
     (z-z_2)-a^2(z-z_4)}\ .
 \end{align}
 Notice that the $z$ transformation is simply a M\"obius
 transformation as expected. The points $(z_2,\chi_2)$ and
 $(z_4,\chi_4)$ are fixed, but in the limit $a\rightarrow0$ all other points approach $(z_4,\chi_4)$ corresponding to the
 collinear limit.

In particular  When $z$ is close to $z_4$ the transformation simplifies to
 \begin{align}
      z-z_4\rightarrow a^2 (z-z_4) + O(z-z_4)^2\qquad \chi -\chi_4 \rightarrow a \,(\chi-\chi_4) + O(z-z_4)\ .
 \end{align}
We see that we are taking a very specific collinear limit, where the
$\chi$'s approach the limit at half the speed that the $z$'s do.

 Thus we have shown that the $(m+1)$-collinear limit  $z_{m+4}, z_{m+2}, \dots
z_{6} \rightarrow z_{4}$ and similarly $\chi_{m+4}, \chi_{m+2}, \dots
\chi_{6} \rightarrow \chi_{4}$ can be implemented (and indeed
explicitly defined) via a family of  superconformal
transformations. Since $R_{m+4}$ is superconformally invariant, the
function is unchanged by the collinear limit, in particular it is
finite and we have $R_{m+4} \rightarrow R_{m+4}$. Thus $R_{m+4}$ is
the $(m+1)$-collinear splitting amplitude.


\section{Symbols and functions at 3-loops}


The conjecture at the centre of the method outlined in \cite{HK3loop} for constructing
MHV amplitudes in special kinematics states that (the logarithms of) the fundamental cross-ratios $u_{ij}$
form the basis for the vector space on which the {\it symbol} of the amplitude is defined.

Fundamental cross-ratios are given by, (cf. Eq.~\eqref{eq:uij})
\begin{align}\label{2eq:uij}
  u_{ij} \,=\, {x_{i, j+1}^2 x_{i+1,j}^2 \over x_{i,j}^2 x_{i+1,j+1}^2} \, =\,
  {\vev{i-1,j+1}\vev{i+1,j-1} \over \vev{i-1,j-1}\vev{i+1,j+1}} \, =\, u_{i-1,i+1;j-1,j+1}
  \, .
\end{align}
For the lowest in $n$ cases, $n=8$ and $n=10$, all non-trivial 2-component cross-ratios
are of the form $u_{i,i+4}$, with $i=1,\ldots,4$ for the octagon, and $i=1,\ldots,10$ for the decagon with the additional
constraint:
\bea \label{constrt8}
n=8\ &:& \qquad 1- u_{i,i+4} = u_{i+2,i+6} \, , \qquad i=1,2 \, \\
n=10\ &:& \qquad 1- u_{i,i+4} = u_{i+2,i+6} \,u_{i-2,i+2} \, , \qquad i=1,\dots,10 .
\eea
At $n=8$ points there are just four fundamental cross-ratios, $u_1$, $u_2$, $v_1$ and $v_2$:
\be \label{uv12}
u_1:= u_{1,5} \quad , \quad
u_2:= u_{2,6} \quad , \quad
u_3:= 1-u_1 := v_1 \quad , \quad
u_4:= 1-u_2 := v_2\, .
\ee

The symbol~\cite{gsv} associates to any (generalised) polylogarithm, a tensor
whose entries are rational functions of the arguments. The rank of the tensor is equal to
the weight of the polylogarithm. For example   $\log x$ has weight 1 and
gives rise to a 1-tensor
\begin{align}\label{eq:11a}
  {\rm Symb}\Big( \log x \Big) = x
\end{align}
whereas the symbol of the classical polylogarithm of weight $n$  is
\begin{align}\label{eq:16a}
  {\rm Symb}\Big(\Li_n(x)\Big) &= - (1-x)\tens \overbrace{x \tens \dots \tens
    x}^{n-1}\ .
\end{align}
The symbol has the properties inherited from the logarithm
\begin{eqnarray}\label{eq:19}
  \dots \tens x\, y \tens \dots &=&\dots \tens x \tens \dots\, +\,  \dots
  \tens y \tens \dots \\
   \dots \tens 1/x \tens \dots &=& -\, \dots \tens x \tens \dots \nonumber
\end{eqnarray}
For the product of functions the symbol
is given by taking the shuffle product of the symbol
of each function
\begin{align}
  {\rm Symb} (f g)= {\rm Symb} (f) \shuffle {\rm Symb}(g)\ .
\end{align}
For example
\begin{align}
  {\rm Symb}( \Li_2(x) \log y) &= \Big(- (1-x) \tens x \Big) \shuffle y
  \nonumber \\
&= -
  (1-x) \tens x \tens y  -
  (1-x) \tens y \tens x  -
  y \tens (1-x) \tens x \, .
\end{align}

In the formalism of \cite{HK3loop} 8-point MHV 3-loop amplitudes have the following structure:
\begin{align}\label{3loopamp2}
  \tilde \cR_8^{(3)}\,=\, \sum_{\sigma,\tau} a_{\sigma\tau} f^{+}_\sigma(u_1)
  f^{+}_\tau(u_2)
\end{align}
where $a_{\sigma\tau}=a_{\tau\sigma}$ are rational coefficients,
and the sum is over the set of functions $f^{+}_\sigma$ with the
properties given in \eqref{eq:9}.
The total polylog weight of $\tilde \cR_8^{(3)}$ must be six which implies that the transcendental weights
of individual functions $f^{+}_\sigma$ can be 2, 4 and 3.
We can now similarly write down the expression for $S_8$,
\bea
S_8^{(3)} (x_2,x_4,x_6,x_8) \,&=&\, \sum_{\sigma,\tau}\,
a_{\sigma\tau} f^{+}_\sigma(u_1) f^{+}_\tau(u_2)\,+\,
b_{\sigma\tau} f^{-}_\sigma(u_1) f^{-}_\tau(u_2) \nonumber\\
& =&\,
\frac{1}{2}\tilde \cR_8^{(3)} \,+\,
T_8^{(3)} (x_2,x_4,x_6,x_8)
\label{S8B}
\eea
with $b_{\sigma\tau}=b_{\tau\sigma}$ and which utilize functions $f^{\pm}_\sigma$
with the property
\be \label{eq:9pm}
f^{\pm}_\sigma(u) \, =\, \pm f^{\pm}_\sigma(v) \quad, \quad v= 1-u\, .
\ee
It can be checked that $T_8$ with correct properties \eqref{T8syms}
indeed arises from the $f^{-}_\sigma(u_1) f^{-}_\tau(u_2)$ combination.
In particular, the transformation of the arguments $(x_2,x_4,x_6,x_8) \leftrightarrow (x_1,x_3,x_5,x_7)$
corresponds in terms of the cross-ratios to $u_2 \leftrightarrow 1-u_2$ with $u_1$
(and $1-u_1$) unchanged. Thus for $T_8^{(3)} (x_2,x_4,x_6,x_8) = b_{\sigma\tau} f^{-}_\sigma(u_1) f^{-}_\tau(u_2) $ ,
for the alternative selection of arguments in $T_8$ we have
$T_8^{(3)} (x_1,x_3,x_5,x_7) =b_{\sigma\tau} f^{-}_\sigma(u_1) f^{-}_\tau(1-u_2)= - b_{\sigma\tau} f^{-}_\sigma(u_1) f^{-}_\tau(u_2) .$
This is in agreement with \eqref{T8syms}.

In~\cite{HK3loop} all possible (symbols and) functions $f^{+}_\sigma(u)$ were listed. It is straightforward to
generalise this construction to functions $f^{\pm}_\sigma$.
For weight-2 there is only one function $f^-$ and one function $f^+$ with properties \eqref{eq:9} or \eqref{eq:9pm},
\be \label{w2-f+}
{\rm weight\, 2:} \quad f^{+}_{\rm weight\,2}(u) = \log(u) \log(v)
\qquad f^{-}_{\rm weight\,2}(u) = \Li_2(u) -\Li_2(v)\, .
\ee
These weight-2 functions are accompanied in \eqref{S8B} by functions $f^{\pm}_\sigma(u)$ of weight-4.
For completeness we list below symbols for all functions $f^{\pm}_\sigma(u)$. They come in two types,
type-a and type-b:
\begin{eqnarray}
   {\rm weight\, 4\,a} &:&
    \begin{array}{l}
{\rm Symb}[f^{\pm}_{a1} ] :=u \tens u \tens
  u \tens v  \pm v \tens v \tens
  v \tens u\\
{\rm Symb}[f^{\pm}_{a2} ] :=u \tens u \tens
  v \tens u \pm v \tens v \tens
  u \tens v\\
{\rm Symb}[f^{\pm}_{a3} ] :=u \tens v \tens
  u \tens u \pm v \tens u \tens
  v \tens v\\
{\rm Symb}[f^{\pm}_{a4} ] :=v \tens u \tens
  u \tens u \pm u \tens v \tens
  v \tens v
  \end{array}
\label{eq:4-A}\\
\nonumber\\
{\rm weight\,4\, b} &:&
    \begin{array}{l}
{\rm Symb}[f^{\pm}_{b1} ] :=u \tens u \tens
  v \tens v   \pm  v \tens v \tens
  u \tens u\\
{\rm Symb}[f^{\pm}_{b2} ]:=u \tens v \tens
  u \tens v \pm v \tens u \tens
  v \tens u\\
{\rm Symb}[f^{\pm}_{b3} ]:=u \tens v \tens
  v \tens u \pm v \tens u \tens
  u \tens v
\end{array}
\label{eq:4-B}
\end{eqnarray}
At the end of Section~{\bf \ref{sec:s_8--contribution}} we explain that
there are no contributions to $\cR_n$ from weight 2 functions
$f^{-}$. Thus there are also no contributions from weight 4 functions
$f^-$ as they would have had to be accompanied by weight 2 functions. 

What remains is to examine the weight-3 functions, known as
type-c. Here we have (cf. \cite{HK3loop}),
\begin{equation}
 {\rm weight\,3\, c} \,:\,
    \begin{array}{l}
 {\rm Symb}[ f^{\pm}_{c1}  ] := u \tens u \tens
  v  \pm  v \tens v \tens u\\
 {\rm Symb}[ f^{\pm}_{c2}  ] :=u \tens v \tens
  u  \pm v \tens u \tens  v \\
 {\rm Symb}[ f^{\pm}_{c3}  ] :=u \tens v \tens v \pm v \tens u \tens u
\end{array}
\label{eq:4cB}
\end{equation}

For the 8-point 3-loop amplitude itself, only the functions $f^+$ appear in  Eq.~\eqref{3loopamp2}.
After imposing the constraint arising from the near-collinear OPE of~\cite{{OPE2}} the final result
of Ref.~\cite{HK3loop} for the octagon at 3-loops
is given by
\begin{align}
  \label{eq:B8-new}
   \tilde \cR_8^{(3)}\,=\,& \log u_1 \log (1-u_1) \Big[ \alpha_1 \,f^+_{a3}(u_2) +\alpha_2
  \,f^+_{a4}(u_2) +\alpha_3 \,f^+_{b2}(u_2)  +\alpha_4 \,f^+_{b3}(u_2) \Big] \nonumber \\
&+\alpha_5  f^+_{c2}(u_1)f_{c2}(u_2)+
\alpha_6 f^+_{c2}(u_1)f^+_{c3}(u_2)+ \alpha_7  f^+_{c3}(u_1)f^+_{c3}(u_2) \nonumber\\
&+ f^+_{c1}(u_1)\Big[ \frac{1}{2} f^+_{c1}(u_2)+
2 f^+_{c2}(u_2)+f^+_{c3}(u_2)\Big]\nonumber\\
& + ( u_1 \leftrightarrow u_2)
\end{align}
with the $f^+_{a}$, $f^+_b$ and $f^+_c$ functions are straightforwardly reconstructed
from their symbols in \eqref{eq:4-A}-\eqref{eq:4cB}  and are listed in Eqs.~(5.15) of Ref.~\cite{HK3loop}.

To fully determine $S_8$ at 3 loops, in addition to $\tilde \cR_8^{(3)}$ we need the contribution
$T_8^{(3)}$ in \eqref{S8B} which comes solely from the $f^-$
functions. Since ultimately  there will be no $f^-$ contributions
at weight 2 (as  shown in Section~{\bf \ref{sec:s_8--contribution}}), the contributions to $T_8^{(3)}$ relevant for $V_n$ can arise only from the weight-3 times weight-3 functions $f^-$
in \eqref{eq:4cB}.
The $f^-$ functions are of the form\footnote{We should note that the third function does not vanish in the collinear limits,
but goes to a constant, $f^-_{c3}(u,v) \to \pm\zeta_3$ when $u$ or $v$ got to 1. This is not a problem, as these constant
terms cancel in the $S$-formula.}:
\begin{eqnarray}
\nonumber
-f^-_{c1}(u,v)&=& \Li_3(u) +\left(\Li_2(v)+\frac{\pi^2}{6}\right) \log(u)+\frac{1}{2}\log(v)\log^2(u) \,-\, (u\leftrightarrow v)
\\\nonumber
f_{c2}(u,v)&=& 2\Li_3(u) +\left(\Li_2(v)-\frac{\pi^2}{6}\right) \log(u)+\log(v)\log^2(u) \,-\, (u\leftrightarrow v)
\\\nonumber
f_{c3}(u,v)&=& \Li_3(u) -\Li_3(v)\, ,
\label{eq:4cB2}
\end{eqnarray}
They give 6 possible combinations,
\begin{equation} \label{eq:T8B}
T_8^{(3)}  \,\,\,\ni\,\,\,
    \begin{array}{l}
f^-_{c1}(u_1,u_3)f^-_{c1}(u_2,u_4)\\
f^-_{c1}(u_1,u_3)f^-_{c2}(u_2,u_4)+f^-_{c2}(u_1,u_3)f^-_{c1}(u_2,u_4)\\
f^-_{c1}(u_1,u_3)f^-_{c3}(u_2,u_4)+f^-_{c3}(u_1,u_3)f^-_{c1}(u_2,u_4)\\
f^-_{c2}(u_1,u_3)f^-_{c2}(u_2,u_4)\\
f^-_{c2}(u_1,u_3)f^-_{c3}(u_2,u_4)+f^-_{c3}(u_1,u_3)f^-_{c2}(u_2,u_4)\\
f^-_{c3}(u_1,u_3)f^-_{c3}(u_2,u_4)
\end{array}
\end{equation}

We now turn our attention to the 10-point amplitude, which was originally obtained in \cite{HK3loop}
in the form given by Eq.~\eqref{eq:14}. The first term on the {\it r.h.s.} gives a particular solution to the
multi-collinear constraints. It is reproduced by the $S_8$ contributions (specifically by the $f^+ f^+$ terms in
\eqref{S8B}. On the other hand, the second term, $V_{10}$ denotes a generic 10-point function which is constrained to
vanish in all triple collinear limits. This collinearly vanishing contribution was constructed in \cite{HK3loop}.

Here, for convenience of the reader, we reproduce the form of $V_{10}$ from \cite{HK3loop}. In order to be able to
uplift the 10-point result to 12 points and all higher points using our general $S$-formula we do not need $S_{12}$
but we need to know that it can be
deconstructed in terms of collinearly vanishing $T_8$ and collinearly vanishing $S_{10}$ contributions.

At 10-points there are 10 fundamental cross-ratios
\be
u_i \, := u_{i,i+4} \, , \qquad i=1,\dots, 10
\ee
which can be divided into 5 parity-even ($u_1, u_3, \dots, u_9$), and 5 parity-odd ross-ratios
($u_2, u_4, \dots, u_{10}$).
It was argued in \cite{HK3loop} that $V_{10}$ is assembled from functions of
even cross-ratios (times functions of odd $u$'s
as follows:
\begin{equation}
  \label{eq:B14}
  f_i(u_{\rm even})f_j(u_{\rm odd}) \ + \ \mbox{cyclic} \ + \ \mbox{parity}\ .
\end{equation}
These functions $f_i$ must
themselves vanish in any collinear limit. To do this they must have
weight-3 and each term must contain 3 consecutive cross-ratios of given parity, eg $u_2,u_4,u_6$.
They are not difficult to find analytically \cite{HK3loop}:
\begin{align}
  \label{eq:B-15}
 f_1(u_2,u_4,u_6) &= \log(u_2) \log(u_4)
 \log(u_6) \nonumber\\
  f_2 (u_2,u_4,u_6)&= \log(u_4)\Big(\Li_2(u_2)
  -\Li_2(1-u_4)+\Li_2(u_6)  -\pi^2/6  \Big) \nonumber\\
f_3(u_2,u_4,u_6,u_8,u_{10})&= \sum_{i=2,4,6,8,10} \Big(\Li_3(u_i)-\Li_3(1-u_i)\Big)-\zeta_3\ .
\end{align}
Here $f_1$ and $f_2$  give 5 independent functions via cyclic permutations of the
arguments, whereas $f_3$ is cyclically symmetric giving only 1
independent function, thus we have 11
functions in total.
These functions are combined together to give a total of 12 independent  weight-6 collinear
vanishing contributions to $V_{10}$:
\begin{align}
  f_1(u_1,u_3,u_5)f_1(u_2,u_4,u_6) \ + \ \mbox{cyclic} \ + \ \mbox{parity}\ \nonumber \\
   f_1(u_1,u_3,u_5)f_1(u_4,u_6,u_8)\ + \ \mbox{cyclic} \ + \ \mbox{parity}\ \nonumber \\
   f_1(u_1,u_3,u_5)f_1(u_6,u_8,u_{10})\ + \ \mbox{cyclic} \ + \ \mbox{parity}\ \nonumber \\
   f_1(u_1,u_3,u_5)f_2(u_2,u_4,u_6)\ + \ \mbox{cyclic} \ + \ \mbox{parity}\ \nonumber \\
   f_1(u_1,u_3,u_5)f_2(u_4,u_6,u_8)\ + \ \mbox{cyclic} \ + \ \mbox{parity}\ \nonumber \\
   f_1(u_1,u_3,u_5)f_2(u_6,u_8,u_{10})\ + \ \mbox{cyclic} \ + \ \mbox{parity}\ \nonumber \\
   f_2(u_1,u_3,u_5)f_2(u_2,u_4,u_6)\ + \ \mbox{cyclic} \ + \ \mbox{parity}\ \nonumber \\
   f_2(u_1,u_3,u_5)f_2(u_4,u_6,u_8)\ + \ \mbox{cyclic} \ + \ \mbox{parity}\ \nonumber \\
   f_2(u_1,u_3,u_5)f_2(u_6,u_8,u_{10})\ + \ \mbox{cyclic} \ + \ \mbox{parity}\ \nonumber \\
  f_1(u_1,u_3,u_5)f_3(u_i^-) \ + \ \mbox{cyclic} \ + \ \mbox{parity}\ \nonumber \\
   f_2(u_1,u_3,u_5)f_3(u_i^-)\ + \ \mbox{cyclic} \ + \ \mbox{parity}\ \nonumber \\
   f_3(u_1,u_3,u_5)f_3(u_i^-)\ + \ \mbox{cyclic} \ + \ \mbox{parity}\
   \label{eq:BlastB}
\end{align}

\end{document}